\documentclass[a4paper,10pt]{article}
\usepackage{amsmath}
\usepackage{amssymb}

\usepackage{sectsty}
\sectionfont{\fontsize{12}{15}\selectfont}

\usepackage{multicol}
\usepackage[export]{adjustbox}
\usepackage{subcaption} 
\usepackage{rotating}

\usepackage[usenames,dvipsnames]{color}
\usepackage{sansmath}
\usepackage[a4paper,margin=1in,footnotesep=2.2\baselineskip]{geometry}
\usepackage{times}
\usepackage{newtxtext,newtxmath}
\usepackage{mathtools}
\usepackage{graphicx}
\usepackage{tikz}
\usepackage[Symbol]{upgreek}
\usepackage{accents}
\usepackage{esint} 
\usepackage{float}
\usepackage{epstopdf}
\usepackage[font=small,labelfont=bf]{caption}
\usepackage{fancyhdr}
\usepackage{changepage}

\usepackage[hang,flushmargin,bottom]{footmisc}

\definecolor{armygreen}{rgb}{0.14, 0.71, 0.15}
\definecolor{darkgreen}{rgb}{0.08, 0.48, 0.18}
\definecolor{darkred}{rgb}{0.86, 0.153, 0.153}
\definecolor{azure}{rgb}{0.0, 0.5, 1.0}
\definecolor{bole}{rgb}{0.82, 0.57, 0.22}

\usepackage[numbers,sort]{natbib}
\usepackage[colorlinks=true,
            linkcolor=blue,
            urlcolor=blue,
            citecolor=blue]{hyperref}
\DeclareCaptionLabelFormat{adja-page}{\hrulefill\\#1 #2 \emph{(\tit{previous page})}}

\newcommand\blfootnote[1]{%
  \begingroup
  \renewcommand\thefootnote{}\footnote{#1}%
  \addtocounter{footnote}{-1}%
  \endgroup
}

\definecolor{armygreen}{rgb}{0.14, 0.71, 0.15}
\definecolor{darkgreen}{rgb}{0.08, 0.48, 0.18}
\definecolor{darkred}{rgb}{0.86, 0.153, 0.153}
\definecolor{azure}{rgb}{0.0, 0.5, 1.0}
\definecolor{bole}{rgb}{0.82, 0.57, 0.22}

\makeatletter
\DeclareSymbolFont{largesymbolsB}{U}{esint}{m}{n}
\re@DeclareMathSymbol{\intop}{\mathop}{largesymbolsB}{'001}
\def\int{\intop\nolimits}
\makeatother

\newcommand*{\dt}[1]{
\accentset{\mbox{\footnotesize\bfseries .}}{#1}}

\def\EatH{Einstein@Home }
\def\SNR{\textsf{SNR}}
\newcommand{\tit}[1]{{\fontfamily{ppl}\selectfont \textit{#1}}}

\def\Hz{$\,$Hz}

\def\ccos{\text{\,\tit{cos}\,}}
\def\csin{\text{\,\tit{sin}\,}}

\def\TwoF{2\mathcal{F}}
\def\Tcoh{\vams{T}{coh}}

\newcommand{\vars}[2]{#1_\mathsf{#2}}
\newcommand{\varr}[2]{#1_\mathrm{#2}}
\newcommand{\vams}[2]{\mathrm{#1}_\mathsf{#2}}
\def\ssb{\textsf{SSB}}
\def\bb{\textsf{BB}}
\def\oone{\textsf{O}1}

\def\sm{\mathrm{M}_\odot}
\def\mn{\mathrm{M}_\mathsf{n}}
\def\mc{\mathrm{M}_\mathsf{c}}

\begin{document}
\fancyhead[L]{\footnotesize\tit{Avneet Singh et al}}
\fancyhead[R]{{\footnotesize \tit{published as} \href{https://journals.aps.org/prd/abstract/10.1103/PhysRevD.100.024058}{\tit{\textbf{Physical Review D}} 100(2):024058}}}
\newpage
\topskip15pt
\begin{center}

\textbf{\large Characterizing the sensitivity of isolated continuous gravitational wave searches to binary orbits}\linebreak

{\small Avneet Singh$^\mathrm{1,\,2,\,3\color{blue}{\dagger}}$, Maria Alessandra Papa$^\mathrm{1,\,2,\,4,\color{blue}{\ddagger}}$\blfootnote{\href{mailto:avneet.singh@aei.mpg.de}{$^\mathrm{{\dagger}}$avneet.singh@aei.mpg.de}; \href{mailto:maria.alessandra.papa@aei.mpg.de}{$^\mathrm{{\ddagger}}$maria.alessandra.papa@aei.mpg.de}}, Vladimir Dergachev$^\mathrm{1}$ \linebreak\linebreak}
{\footnotesize $^1$ Max-Planck-Institut f{\"u}r Gravitationphysik, Callinstra{$\upbeta$}e 38, 30167, Hannover\\
$^2$ Leibniz Universit{\"a}t Hannover, Welfengarten 1, 30167, Hannover\\
$^3$ The Geophysical Institute, Bjerknes Centre for Climate Research, University of Bergen, Bergen 5007, Norway\\
$^4$ University of Wisconsin-Milwaukee, Milwaukee, Wisconsin 53201, USA}

\setcounter{footnote}{0}

\begin{abstract}
Broadband all-sky searches for continuous gravitational waves (CW) from isolated neutron stars typically yield a number of significant candidates that are discarded through hierarchical semi-coherent and coherent follow-up searches. However, these searches target purely isolated sources and do not take into account possible phase modulations to the signals due to companion objects in binary orbits. If the source object is in a binary orbit, the signal-to-noise ratio ({\SNR}) of a signal will diminish due to the mismatch between the signal waveform and the template waveform used in the search. In this paper, we investigate the sensitivity of CW searches to signals from sources in binary systems over an exhaustive range of binary orbits. We use Monte-Carlo simulations to constrain the ranges of binary orbital parameters that leave a CW signal from a source in a binary orbit visible in a search for signals from an isolated source (\tit{isolated searches}), and consequently tighten the parameter space explorable with dedicated searches for signals from sources in binary systems (\tit{binary searches}).
\end{abstract}
\end{center}

\begin{multicols}{2}
\section{Introduction}
\label{sec:intro}

In recent years, there has been a significant increase in efforts to search for continuous gravitational waves (CW) from neutron stars. The CW sources targeted by these searches include unknown isolated neutron stars \citep{S6BucketStage0,S6FU,O1AS20-100, O1MultiPipeline,O2ASLIGO,Dergachev:2019pgs,Dergachev:2019wqa}, known isolated neutron stars \citep{Authors:2019ztc,Abbott:2018qee,S6CasA,Ming:2019vb}, unknown neutron stars in binary systems \citep{S6binary} and known neutron stars in binary systems such as Scorpius X-1 \citep{O1ScoX1,S6ScoX1}.  

In {\citep{O1AS20-100}}, several thousand candidates were found in the initial semi-coherent stages of the search. While a large majority of the candidates were ruled out in the hierarchical semi-coherent and fully-coherent follow-up and vetoing stages \citep{DMOff}, the few remaining survivors (four) were eventually discarded through searches in a different data-set. The new data-set spanned a period of time over an year apart from the time period spanned by the original search. In such a scenario, even a small mismatch between the signal and template waveform during the original search could evolve into a much larger mismatch in a search on a newer/different data-set, resulting in a significant loss of {\SNR} and consequently leading to false dismissal of a legitimate signal. This could happen if the signal is emitted by a source in a binary system whereas the search specifically targets emission from isolated objects. In some cases, in the presence of a binary companion object, a CW signal may also appear to turn on and off and on again in subsequent searches as a function of the binary orbital phase of the source object, thus exhibiting a dynamic transient nature unique to binary systems\footnote{This transient nature of the CW signal derives from a coupling between the orbital period of the binary system and the time lapsed between two searches.} on top of the aforementioned mismatch evolution. Since the most extensive broadband all-sky searches can only afford to look for emission from an isolated source, the issue of a candidate signal from a source in a binary system identified in one data-set getting discarded by a subsequent follow-up search on a different data set is very relevant. On the other hand, it is not good practice to evolve the signal model based on \tit{a posteriori} information gained from the data in searches for very weak signals; in these cases, it is usually very hard to formulate an unbiased interpretation of the significance of one's findings. One must instead understand the response of all-sky \tit{isolated searches} (referring to CW searches targeting isolated sources) to signals from objects in generic binary orbits and, \tit{a priori} prepare a sensible follow-up strategy.

To address this, we study the effects of possible binary companion objects on the detectability of a CW source. We narrow down on the ranges of binary orbital parameters that leave a modulated signal detectable in an isolated search by, \tit{a}) analytically simplifying the problem and calculating the expected drift in intrinsic signal parameters and the loss of {\SNR} in the presence of a companion object \tit{aka} `the binary hypothesis', and \tit{b}) using Monte-Carlo simulations to confirm the detectability as a function of the binary orbital parameters. The results of the Monte-Carlo simulations also enable us to determine the drift in the intrinsic parameters more precisely under the binary hypothesis since they fold in all the subtle effects of an isolated search that are lost in analytic simplifications. In these reduced ranges of the orbital and intrinsic parameters, targeted binary searches could perform more sensitive searches for sources in binary orbits \citep{ScoX1MDC,Dhurandhar:2000sd,binaryprix}.

\section{A typical binary system}
\label{sec:system}
\begin{figure}[H]
\centering\includegraphics[width=78mm]{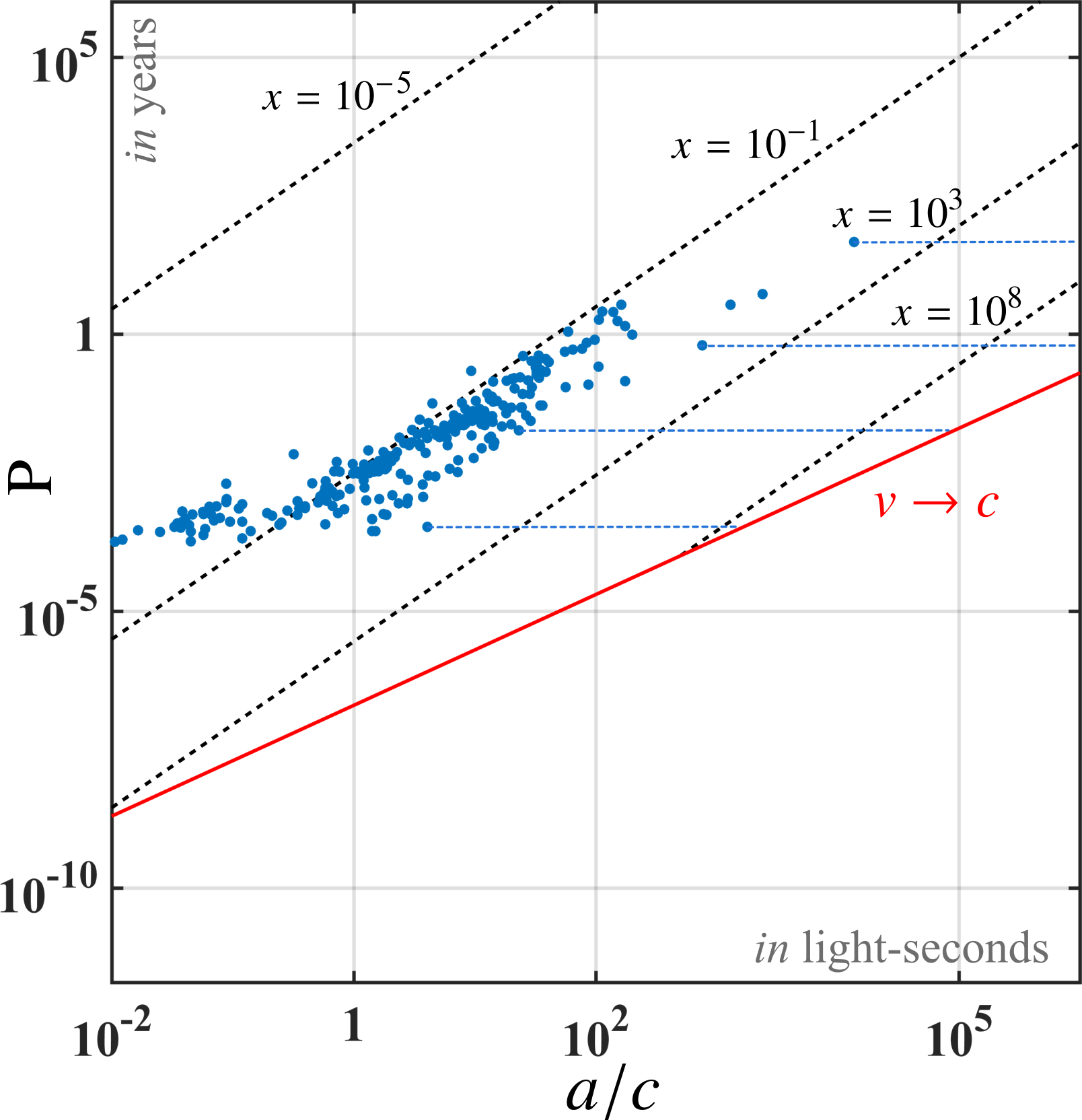}
\caption{{\small The orbital period $\vams{P}{b}$ (in years) versus semi-major axis $a$ (in light-second units) for binary systems as function of fixed mass ratios $x$ (dashed lines). The red line encodes the physical limit where the mean orbital speed $v$ of the source tends to the speed of light, i.e. $v \sim 2\uppi\,a/\vams{P}{b} \rightarrow c$. The over-plotted blue points represent all known companion objects in a binary orbit with a neutron star, taken from the ATNF catalogue \citep{ATNF}. Note that the ATNF data only provides the \tit{projected} semi-major axis $\vars{a}{p}$ ($= a \csin i$) and not the physical semi-major axis $a$ since most astronomical observations are only sensitive to the projection of the binary orbit in the absence of polarization information. Thus, to show all possible values of $a$ corresponding to a given $\vars{a}{p}$, i.e. $\vars{a}{p} \leq a \leq c\vams{P}{b}/2\uppi$, the blue points must be interpreted as lines extending to the limit $v\rightarrow c$, as shown for a few selected objects. This interpretation shall be assumed throughout this paper.}}
\label{fig:system}
\end{figure}

To begin with, consider a system consisting of a source object in a binary orbit with an unknown companion object. We assume the source object, i.e. a neutron star, to have a mass ($\mn$) of 1.4 $\sm$, while the mass of the unknown companion object ($\mc$) is allowed to vary freely. The mass ratio ($\mc/\mn$) is denoted by $x$ and the semi-major axis of the source object's orbit is denoted by $a$. In the non-relativistic Newtonian limit, the period ($\vams{P}{b}$) of the source orbit is given by Kepler's laws as
\begin{equation}
\vams{P}{b} = \frac{2\uppi}{\sqrt{\mathrm{G}\mn}}\frac{1+x}{x^{3/2}}a^{3/2}.\label{eq:period}
\end{equation}
{\noindent}In figure \ref{fig:system}, we plot the relation \eqref{eq:period}, along with all known companion objects in a binary system with a neutron star taken from the ATNF catalogue \citep{ATNF}.

\section{The phase model}
\label{sec:phmodel}
\begin{figure}[H]
\centering\includegraphics[width=78mm]{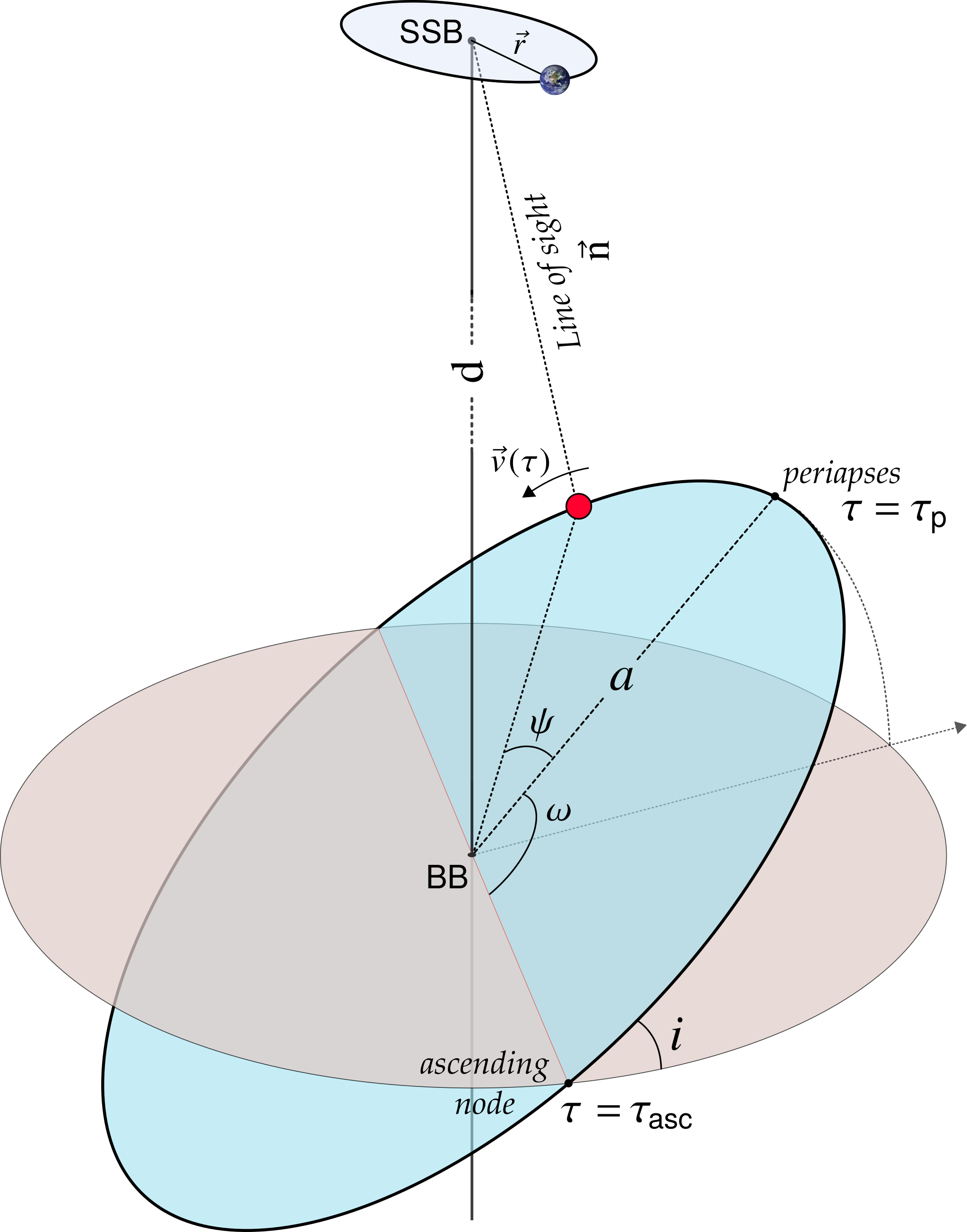}\vspace{20pt}
\centering\includegraphics[width=78mm]{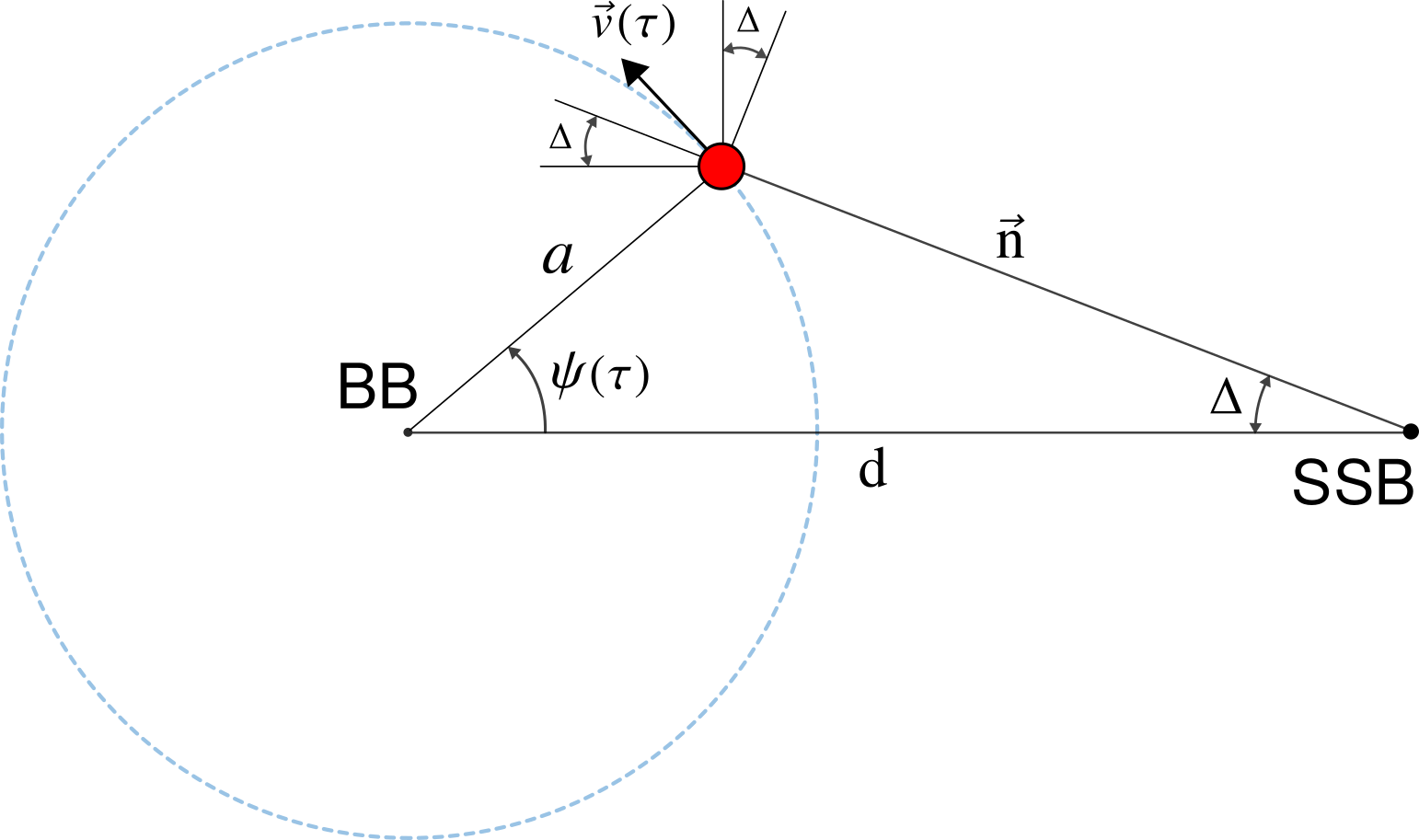}
\caption{{\small \tit{Top}: Binary orbit with its orbital parameters. \tit{Bottom}: Projection in two dimensions, simplified assuming a circular orbit with $i = \uppi/2$, for ease of geometrical interpretation. Typically, $\Delta \rightarrow 0$ since $a \ll \mathrm{d}/c$.}}
\label{fig:orbit}
\end{figure}

Consider that the source object is emitting long-lasting and nearly monochromatic continuous gravitational waves. This CW emission may be the result of a rigid asymmetry in the star's shape (\tit{aka} non-zero ellipticity), fluid oscillations such as $r$-modes, precession etc \citep{Sh}. Irrespective of the emission model, the resulting phase model for the emitted wavefront in the source frame (denoted by time $\tau$) is given by
\begin{equation}
\phi(\tau)=2\uppi\bigg[f(\tau-\vars{\tau}{ref})+\frac{1}{2}\,\dt{f}(\tau-\vars{\tau}{ref})^2+ \cdots\bigg],\label{eq:phin}
\end{equation}
where,  $\vars{\tau}{ref}$ is the reference time, $f = \partial_\tau\phi|_{\tau=\vars{\tau}{ref}}$ is the CW frequency and $\dt{f} = \partial^2_\tau \phi|_{\tau=\vars{\tau}{ref}}$ is the first-order spin-down calculated at $\tau=\vars{\tau}{ref}$. The wavefront denoted by \eqref{eq:phin} propagates through space to the detector location on Earth at a different time $t$ which is related to the time of emission $\tau$ by
\begin{equation}
\tau=t + \frac{\vec{r}(t)\cdot\vec{\mathrm{n}}}{c} - \frac{\mathrm{d}}{c}-\frac{\mathrm{R}(\tau)}{c},\label{eq:tarr}
\end{equation}
where, $\mathrm{d}$ is the distance between the binary barycenter (\bb) and the solar-system barycenter ({\ssb}), $\mathrm{R}(\tau)$ is the projected distance of the emitting neutron star from the {\bb} along the light of sight, $\vec{r}$ is the vector pointing from of the {\ssb} to the location of the detector, and $\vec{\mathrm{n}}$ is the unit vector pointing from the {\ssb} toward the source\footnote{Note that the relativistic wave propagation effects, e.g. Shapiro delay, have been ignored since they are of negligible magnitude \citep{binaryprix, releff1, releff2}.} (see figure \ref{fig:orbit}). We may assume that the detectors are located at the {\ssb}, hence ignoring the second term, such that
\begin{equation}
\tau = t - \frac{\mathrm{d}}{c}-\frac{\mathrm{R}(\tau)}{c}.\label{eq:nred}
\end{equation}
We allow this simplifying assumption since the relative motion of the detector on Earth with respect to the {\ssb}, and in general to any {\bb}, is accounted for in an isolated all-sky CW search for any given sky position. Hence, we only need to evaluate the \tit{residual} modulations due to the motion of the source around its companion object and the impact of such a motion on the detectability of the CW signal and on parameter estimation. Accordingly, we will use the terms `Doppler modulation' and `phase mismatch' referring to specifically the \tit{residual} Doppler modulation and phase mismatch for brevity throughout the rest of the paper.

We note that the term ${\mathrm{d}}/{c}$, which depends on the distance of a specific source from the {\ssb}, is not accounted for in an all-sky CW search which probes only directions and \tit{not} distances. This does not affect the sensitivity of the search since its effect is a constant difference between $\tau$ and $t$, which is equivalent to a re-labeling of the intrinsic signal parameters.

The orbital motion of an object in binary orbit is generally expressed in terms of the following parameters: the eccentricity ($e$), the argument of periapsis ($\omega$), the semi-major axis ($a$), the time of ascending node ($\vars{\tau}{asc}$) and the orbital period ($\vams{P}{b}$). In equation \eqref{eq:nred}, the last term containing the binary orbit is then written in terms of the mean and eccentric anomaly -- $\upnu(\tau)$ and $\mathrm{E}(\tau)$ respectively, as
\begin{equation}
\begin{multlined}
\frac{\mathrm{R}(\tau)}{c} = \vars{a}{p}\big[1 - e \ccos \mathrm{E}(\tau)\big] \csin[\upnu(\tau) + \omega],\label{eq:R}
\end{multlined}
\end{equation}
where, $\vars{a}{p}=a\csin i/c$ (see figure \ref{fig:orbit}), and the two anomalies are related by the implicit relation $\upnu(\tau) = \mathrm{E}(\tau) - e \csin \mathrm{E}(\tau)$. By combining \eqref{eq:phin}, \eqref{eq:nred} and \eqref{eq:R}, we can fully describe the time-varying phase model for the emitted CW signal as a function of the arrival time $t$. 
{\noindent}The effect of the $\mathrm{R}(\tau)$ term is a modulation in the phase of the CW measured by an observer at rest with respect to the {\ssb}, due to the binary orbit. In an isolated search, this modulation is not accounted for, and it generally leads to a loss of sensitivity and the maximum of the detection statistic to shift in the observed signal parameters ($f,\dt{f},\alpha,\delta$). Our interest lies is in estimating the range of binary orbital parameters where the loss in sensitivity is relatively small, and in calculating the shift in the observed signal parameters. In order to do this, we will consider the case of nearly circular orbits, and evaluate the Doppler shift in frequency averaged over observation time) and the induced average spin-down due to the time-variation in Doppler modulation. We expect these two quantities to mimic the expected shift in the intrinsic signal parameters in an isolated search.

\section{Testing the nearly circular limit}
\label{sec:lowe}
In order to perform a characterization of the response of an isolated search to binary signals, we consider the zero-eccentricity limit ($e\rightarrow0$); this is not a constraining assumption since binary orbits are expected to circularize over time and such systems are the primary targets of CW searches and about 75\% binary pulsars in the ATNF catalogue have $e < 0.01$ \citep{ATNF}. In this case, the eccentric and mean anomaly depend on the orbital phase according to $\upnu(\tau) = \mathrm{E}(\tau) = \psi(\tau) = \Omega(\tau - \vars{\tau}{p})$. We express $\vars{\tau}{p}$ in terms of the time of ascending node by $\vars{\tau}{asc}=\vars{\tau}{p} - \omega/\Omega$, where $\Omega$ is the mean orbital frequency ($=2\uppi/\vams{P}{b}$). In this limit, $\mathrm{R}(\tau)$ is given by
\begin{equation}
\begin{multlined}
\mathrm{R}(\tau)/c = \vars{a}{p}\csin \psi(\tau) = \vars{a}{p}\csin [\Omega(\tau - \vars{\tau}{p})].\label{eq:phi0e}
\end{multlined}
\end{equation}
The resulting phase model is then written by combining \eqref{eq:phi0e}, \eqref{eq:nred} and \eqref{eq:phin}, and it depends only on $\vars{a}{p}$, $\vams{P}{b}$ and $\vars{\tau}{p}$ ($\equiv \vars{\tau}{asc}$), freeing us from $e$ and $\omega$. The observed instantaneous frequency and spin-down of the emitted signal at any time $t$ are calculated by computing the first and second time-derivative of the phase model, i.e. $\partial_t\phi$ and $\partial_{tt}\phi$ respectively.

\subsubsection*{\centering \tit{Doppler shift \& induced spin-down}}
\label{sec:doppler}
The main contribution to the mismatch between a binary signal and an isolated signal stems from the phase discrepancy between the two signals. In the limit of orbital periods longer than the coherent observation time $\vams{T}{coh}$ of the CW search, one expects the mismatch to be described by the average orbital Doppler modulation, and this is the hypothesis that we want to test. We also expect that as the orbital period becomes comparable to the coherent observation time, the mismatch should diverge from the expectation derived from average orbital Doppler modulation.

The frequency $f(t)$ measured at time $t$ by an observer moving with a velocity $\vec{v}(t)$ with respect to a source in the direction $\vec{\mathrm{n}}$ (for example, figure \ref{fig:orbit} bottom panel) and with an intrinsic frequency $\varr{f}{o}$ is 

\begin{figure}[H]
\centering\includegraphics[width=78mm]{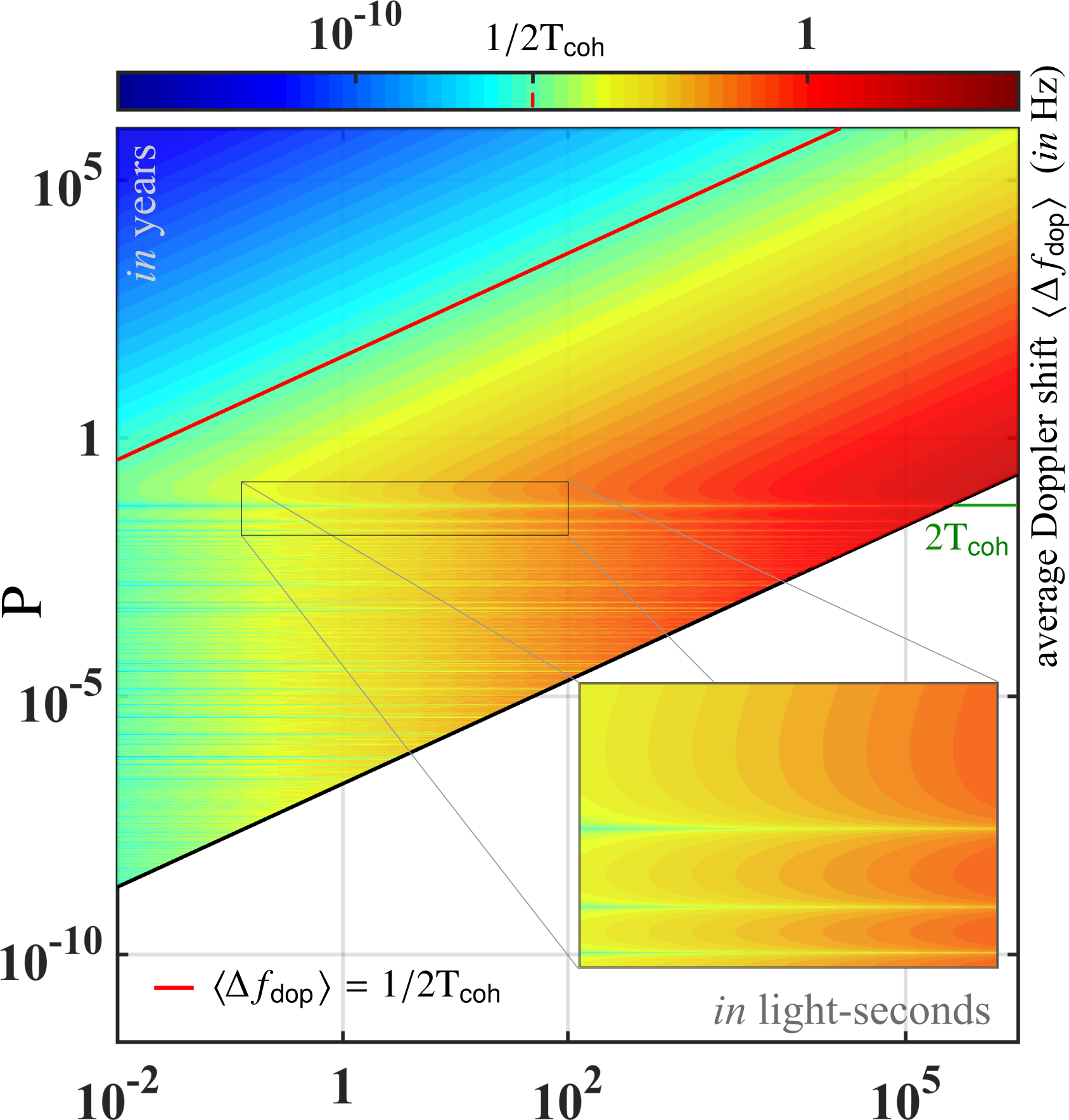}
\centering\includegraphics[width=78mm]{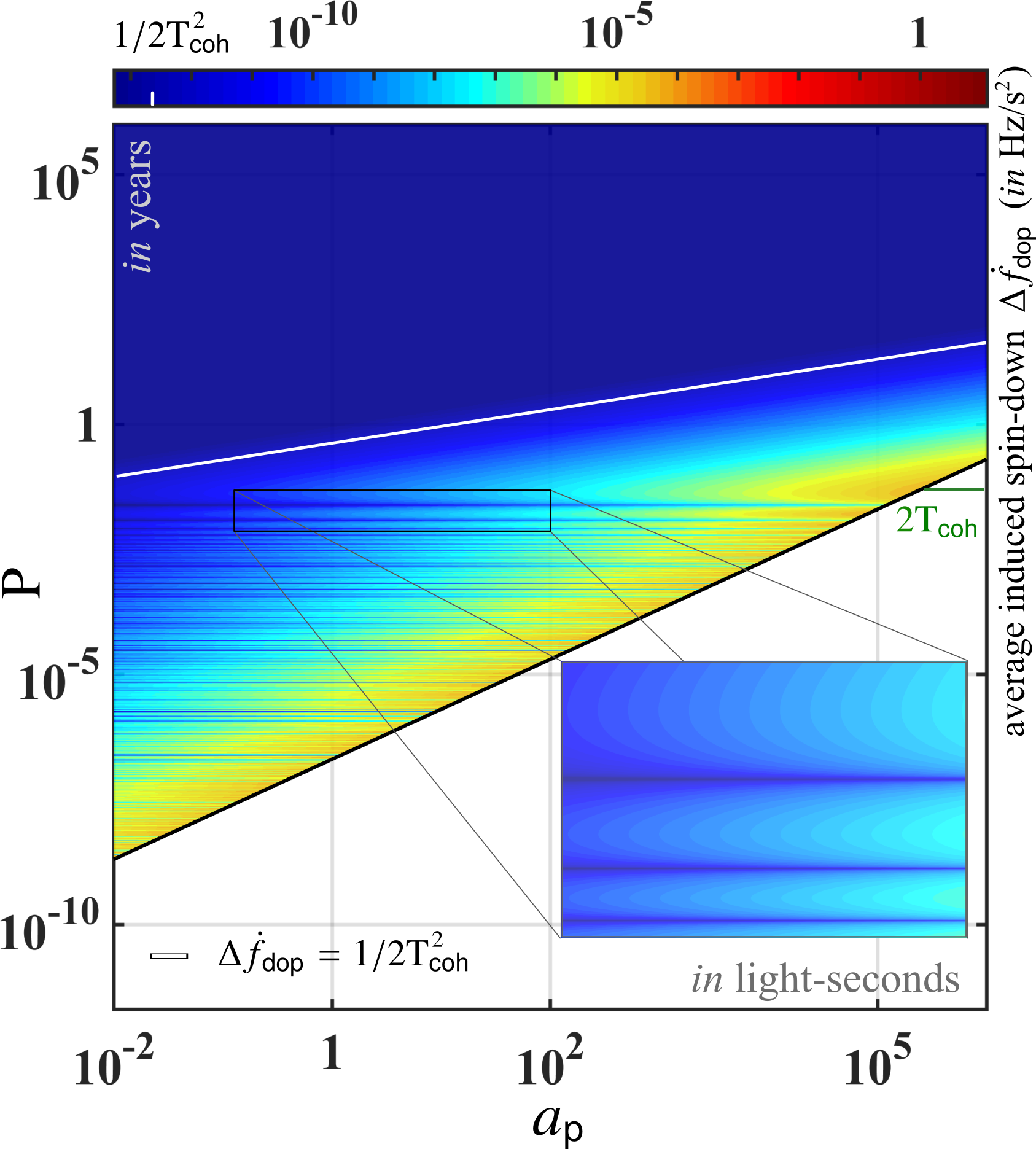}
\caption{{\small Doppler shift and induced spin-down as a function of $\vars{a}{p}$ ($= a \csin i/c$) and $\vams{P}{b}$ calculated over 210 hours of coherent observation time ($\vams{T}{coh}$). In a typical CW search, the frequency and spin-down resolution are given by $\sim 1/2\vams{T}{coh}$ and $\sim 1/2\vams{T}{coh}^2$ respectively; these contours are marked with a red line in the top panel and a white line in the bottom panel. The regions above these lines are expected to give rise to unresolvable shifts in the intrinsic parameters while the regions below show increasingly larger shifts as one moves away from these lines. We also see that as the orbital period becomes comparable to $\vams{T}{coh}$, oscillations appear as a function of $\vars{\psi}{coh}$ described by \eqref{eq:c3f}.}}
\label{fig:doppler}
\end{figure}

\begin{equation}
\begin{gathered}
\frac{f(t) - \varr{f}{o}}{\varr{f}{o}} = \frac{\Delta\vars{f}{dop}(t)}{\varr{f}{o}} = \sqrt{\frac{1+\gamma}{1-\gamma}} - 1,\;\;\;\text{where,}\\
\gamma(t) = \frac{1}{c}\,\vec{v}(t)\cdot\vec{\mathrm{n}}.
\end{gathered}\label{eq:doppler}
\end{equation}

{\noindent}The intrinsic frequency $\varr{f}{o}$ is the frequency that would be measured by an observer at rest with respect to the source at a reference time $\varr{\tau}{o}$. We denote the corresponding detector-time by $\varr{t}{o}$. If the source is very far from us, i.e. $\Delta \rightarrow 0$ in figure \ref{fig:orbit}, and in a binary circular orbit around a companion with $v \ll c$, the average Doppler shift over an observation time $(t_1,t_2)$, denoted by $\langle\Delta\vars{f}{dop}\rangle$, is 
\begin{equation}
\begin{gathered}
\frac{\langle\Delta\vars{f}{dop}\rangle}{f} = \frac{1}{f\,|t_2-t_1|} \int_{t_1}^{t_2} \Delta\vars{f}{dop}(t^\prime)\,\mathrm{d}t^\prime \xrightarrow{\displaystyle{v}\ll c} \\ \frac{1}{|t_2-t_1|}\int_{t_1}^{t_2}\gamma(t^\prime)\,\mathrm{d}t^\prime = \frac{\csin i}{|t_2-t_1|}\frac{v}{c}\int_{t_1}^{t_2}\ccos \psi(t^\prime)\,\mathrm{d}t^\prime,
\end{gathered}\label{eq:dopred}
\end{equation}
The peak magnitude of the Doppler shift is calculated by simply setting $\gamma = {|v|}/c$ in \eqref{eq:doppler}, i.e. when $\ccos \psi(t^\prime)=1$.

The Doppler shift in \eqref{eq:dopred} is a time-dependent quantity. Thus, one can define a first-order induced spin-down in the interval $(t_1,t_2)$ associated with the Doppler shift as
\begin{equation}
\Delta\vars{\dt{f}}{dop} = \frac{f(t_2) - f(t_1)}{|t_2-t_1|} = \frac{\Delta\vars{f}{dop}(t_2) - \Delta\vars{f}{dop}(t_1)}{|t_2-t_1|}.\label{eq:fdotdop}
\end{equation}
One could imagine that search sensitivity of an isolated search to a binary signal might be regained for an isolated signal waveform (template) with a spin-down value that can `add-up to' the orbital Doppler modulation described by \eqref{eq:dopred}.

Expressions \eqref{eq:dopred} and \eqref{eq:fdotdop} are simplified for generic values of $\vams{P}{b}$ and $|t_2-t_1|=\vams{T}{coh}$ to
\begin{equation}
\begin{gathered}
{\langle\Delta\vars{f}{dop}\rangle} = \frac{2\uppi f\vars{a}{p}}{\vams{P}{b}} \frac{\csin \vars{\psi}{coh}}{\vars{\psi}{coh}},\;\;\;\\
\Delta\vars{\dt{f}}{dop} = \frac{4\uppi^2 f\vars{a}{p}}{\mathrm{P}^2_\mathsf{b}} \frac{\ccos \vars{\psi}{coh} - 1}{\vars{\psi}{coh}},
\end{gathered}\label{eq:c1f}
\end{equation}
with $\vars{\psi}{coh} \equiv \Omega(t_2 - t_1) = \Omega\vams{T}{coh}$ being the orbital phase subtended by the source object in time $\vams{T}{coh}$ starting at $t_1 = t(\vars{\tau}{p})$.

$\langle\Delta\vars{f}{dop}\rangle$ and $\Delta\vars{\dt{f}}{dop}$ are plotted in figure \ref{fig:doppler} as a function of $\vars{a}{p}$ and $\vams{P}{b}$, for an observation time $|t_2-t_1|$ of 210 hours and assuming the start-time to be $t_1 = t(\vars{\tau}{p})$. These two quantities encode the broad features of the response of an isolated search to a binary signal. In order to understand the different transitions in figure \ref{fig:doppler}, we dissociate the expressions \eqref{eq:dopred} and \eqref{eq:fdotdop} into two constituent cases: \tit{a}) long-period limit,  $\vams{P}{b} \gg \vams{T}{coh}$, and \tit{b}) short-period limit,  $\vams{P}{b} \ll \vams{T}{coh}$.

\begin{center}
{{\tit{a}) \tit{long-period limit}}}
\end{center}
In the limit $\vams{P}{b} \gg \vams{T}{coh}$, \eqref{eq:c1f} reduces to
\begin{equation}
\begin{gathered}
{\langle\Delta\vars{f}{dop}\rangle} \sim \frac{2\uppi f\vars{a}{p}}{\vams{P}{b}},\;\;\;\\
\Delta\vars{\dt{f}}{dop} \sim -\frac{\uppi^2 f\vars{a}{p}}{\mathrm{P}^2_\mathsf{b}}\vars{\psi}{coh} = -\frac{2\uppi^3 f\vars{a}{p}}{\mathrm{P}^3_\mathsf{b}}\vams{T}{coh},
\end{gathered}\label{eq:c2f}
\end{equation}
since ${\csin \vars{\psi}{coh}}/{\vars{\psi}{coh}}\rightarrow 1$. On log scale, \eqref{eq:c2f} represents straight lines of slope 1 and 1/3 respectively, as seen in figure \ref{fig:doppler}. In this limit, the resultant phase model closely mimics a standard isolated CW signal and we expect the observed shift in frequency and spin-down to be well-described by the analytic expression \eqref{eq:c2f}. 

\begin{center}
{{\tit{b}) \tit{short-period limit}}}
\end{center}
When the orbital period becomes comparable to or shorter than the coherent observation time (i.e. $\vams{P}{b} \lesssim \vams{T}{coh}$), the resulting phase model diverges from a standard CW signal template and the observed shifts cannot be accurately derived from \eqref{eq:c1f}. In the limit $\vams{P}{b} \ll \vams{T}{coh}$, \eqref{eq:c1f} equates to
\begin{equation}
\begin{gathered}
{\langle\Delta\vars{f}{dop}\rangle} = \frac{f\vars{a}{p}}{\vams{T}{coh}}\csin \vars{\psi}{coh},\;\;\;\text{and,}\\
\Delta\vars{\dt{f}}{dop} = -\frac{2\uppi f\vars{a}{p}}{\vams{T}{coh}\vams{P}{b}}\ccos \vars{\psi}{coh}-1,
\end{gathered}\label{eq:c3f}
\end{equation}
respectively. We note that \eqref{eq:c3f} shows periodicity as a function of $\vars{\psi}{coh}$. In figure \ref{fig:doppler}, we see this periodic nature clearly, notably in the zoomed-in inset windows. We also find that these periodic oscillations get tighter as the orbital period becomes shorter, and in the extreme limit, the result is effectively randomised. In this limit, the results quoted in \eqref{eq:c3f} and shown in figure \ref{fig:doppler} cannot be trusted since the resultant phase model is badly approximated by a standard isolated CW signal template. 
\begin{figure}[H]
\centering\includegraphics[width=78mm]{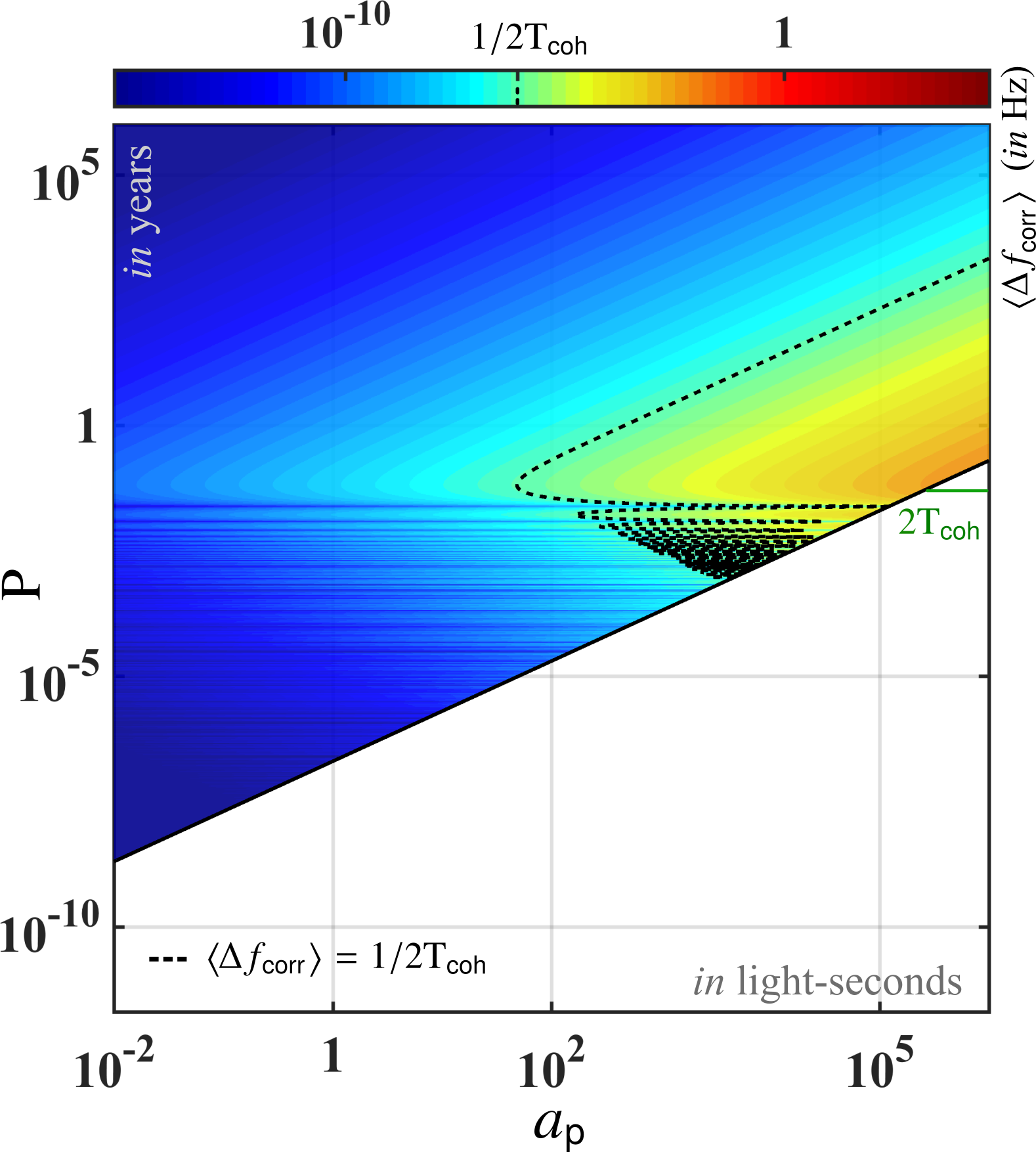}
\caption{{\small Correction in frequency due to the ${\mathrm{R}(\tau)}/{c}$ term in the timing model calculated for $\vams{T}{coh} = $ 210 hours.}}
\label{fig:corr}
\end{figure}
\subsubsection*{\centering \tit{Timing correction(s)}}
\label{sec:corr}
In equation \eqref{eq:nred} of section \ref{sec:phmodel}, we recall the presence of time-dependent term ${\mathrm{R}(\tau)}/{c}$ in the timing model. The effect of this term is similar to that of $\mathrm{d}/c$ term, in that it amounts to a relabeling of the intrinsic parameters as discussed before, but with the added complexity of time-dependence. The overall effect of this term is an additional shift of the observed signal frequency of magnitude $\dt{f}{\mathrm{R}(\tau)}/{c}$, similar to the frequency shift $\dt{f}\mathrm{d}/c$ due to the term $\mathrm{d}/c$. The average frequency shift due to this time-varying term is given by
\begin{equation}
\begin{gathered}
{\Delta\vars{f}{corr}} = {2\uppi\,\vars{a}{p}\dt{f}}\frac{\ccos \vars{\psi}{coh} - 1}{\vars{\psi}{coh}}.
\end{gathered}\label{eq:fcorr}
\end{equation}
In typical CW searches, the explored magnitudes of the $\dt{f}$ parameter lie in the range $10^{-11} - 10^{-8}\,\text{Hz/s}$. The relative contribution of this correction ($\propto\dt{f}\,|\ccos\vars{\psi}{coh} - 1|$) is either very small compared to the Doppler shift ($\propto f\csin\vars{\psi}{coh}/\vams{P}{b}$), or comparable but unresolvable, for the values of $f$, $\dt{f}$, $\Tcoh$ and $\vams{P}{b}$ considered in our simulations (as shown in figure \ref{fig:corr}), so we ignore this timing correction in further discussions.

\subsubsection*{\centering\tit{Impact of }T{$\vams{}{coh}$}}
\label{sec:tcoh}
\begin{figure}[H]
\centering\includegraphics[width=78mm]{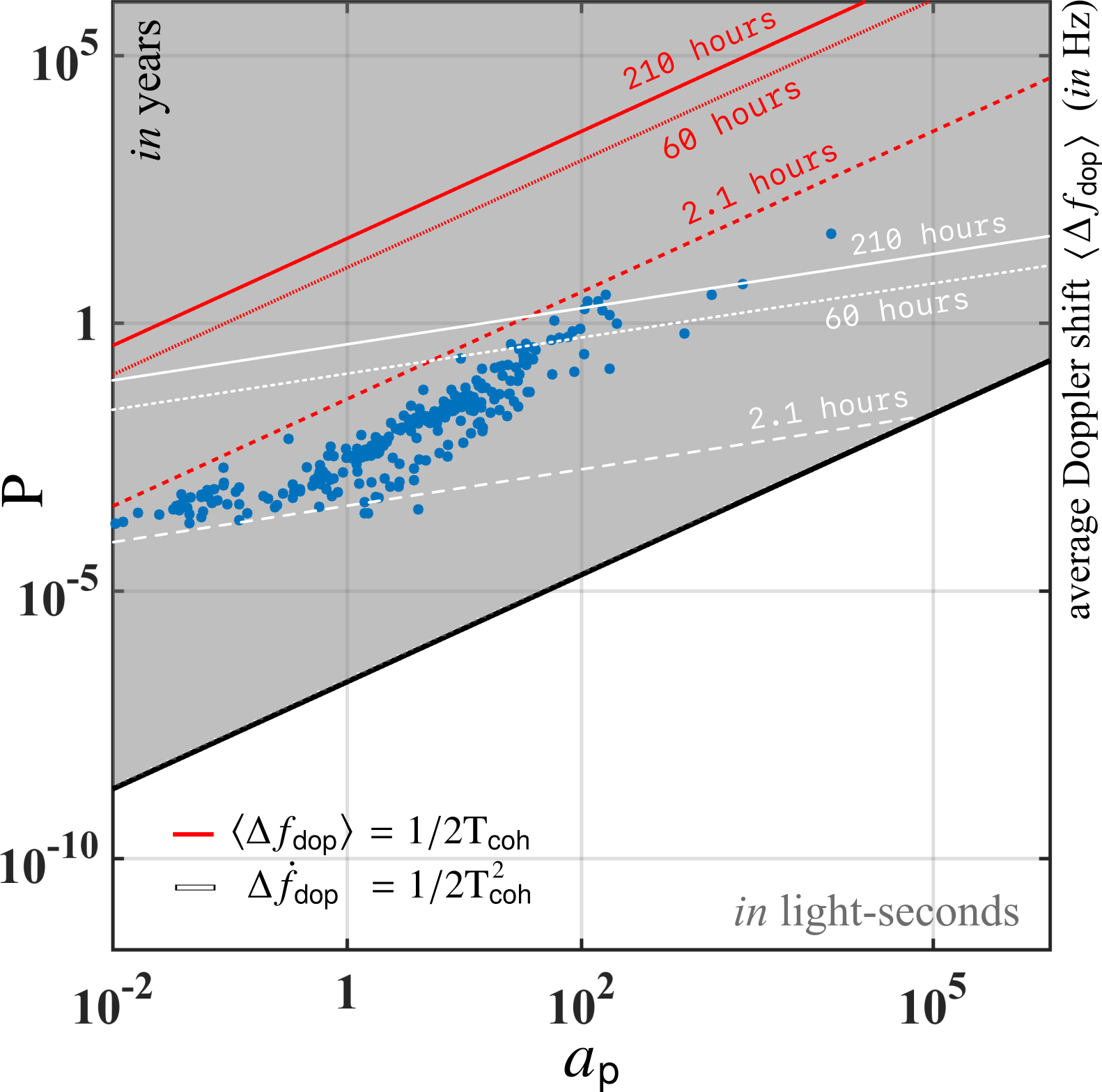}
\flushleft\includegraphics[width=74.5mm]{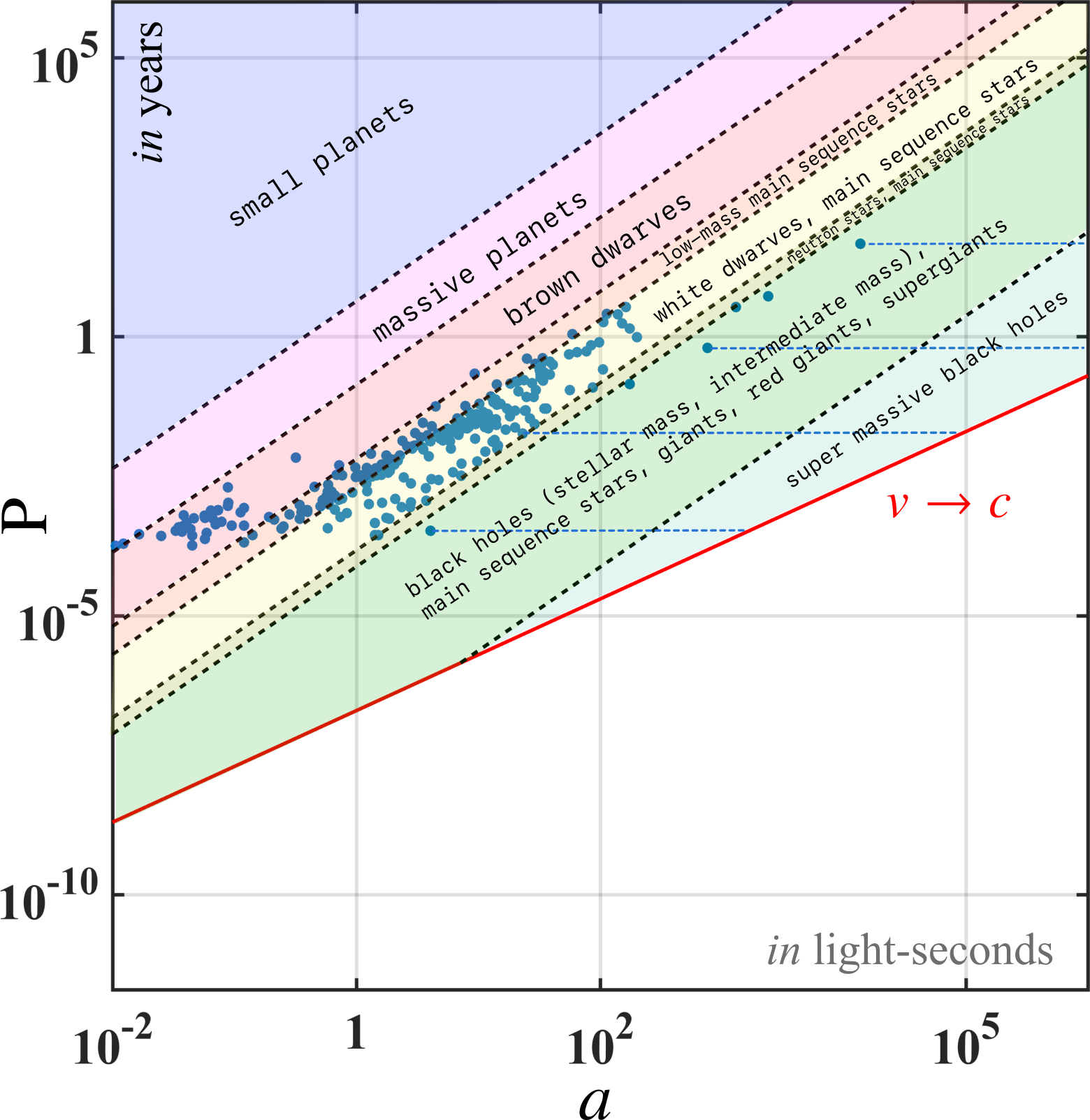}
\caption{\tit{Top panel}: Lines of constant average residual Doppler shift and average residual spin-down equal to the natural resolutions of searches with three different coherent time baselines $\vams{T}{coh}$, ranging from 2.1 to 210 hours. \tit{Bottom panel}: Populations of different objects identified in the $a-\vams{P}{b}$ plane.}
\label{fig:tcoh}
\end{figure}

In the previous section, we found that the expected shifts in the intrinsic values of frequency and spin-down depend on the choice of coherent observation time $\vams{T}{coh}$. This dependence is summarized in figure \ref{fig:tcoh} (top panel) for three different choices of $\vams{T}{coh}$ -- 2.1 hours, 60 hours and 210 hours. We find that with decreasing $\vams{T}{coh}$, increasingly larger regions in the $\vars{a}{p}-\vams{P}{b}$ give rise to shifts in the observed intrinsic frequency and spin-down parameters which are smaller than the natural resolutions ($1/{2\vams{T}{coh}}$ and $1/{2\mathrm{T}^2_\mathsf{coh}}$ in $f$ and $\dt{f}$ respectively). This essentially means that the modulation effects of the binary companion object on a CW source are tracked more efficiently by an isolated search when the coherent time baselines are shorter in length. This is expected since the deviations between signal and template waveforms have less time to accumulate over shorter coherent lengths. The drawback, however, is that the total {\SNR} also has less time to accumulate, thus reducing the overall sensitivity of such searches. We will observe this sensitivity loss in the next section.

In figure \ref{fig:tcoh}, we also overlay the known companion objects in binary orbit with known neutron stars showing the physical ranges of $\vars{a}{p}$ and $\vams{P}{b}$ spanned by known systems; this data is taken from the ATNF catalogue \citep{ATNF}. For a more general perspective, the bottom panel in figure \ref{fig:tcoh} shows division into different categories classified by object types, from small planets to super-massive black holes. 

\section{Response of an isolated search} 
\label{sec:reslowe}
In the previous section, we presented a simplified model to estimate the apparent shifts in signal parameters $f$ and $\dt{f}$ occurring when a binary signal is identified with an isolated search. In this section, we attempt to corroborate these estimates with direct search simulations.

\subsubsection*{\centering \tit{Discretisation of binary parameter space}}
\label{sec:pixels}
\begin{figure}[H]
\centering\includegraphics[width=78mm]{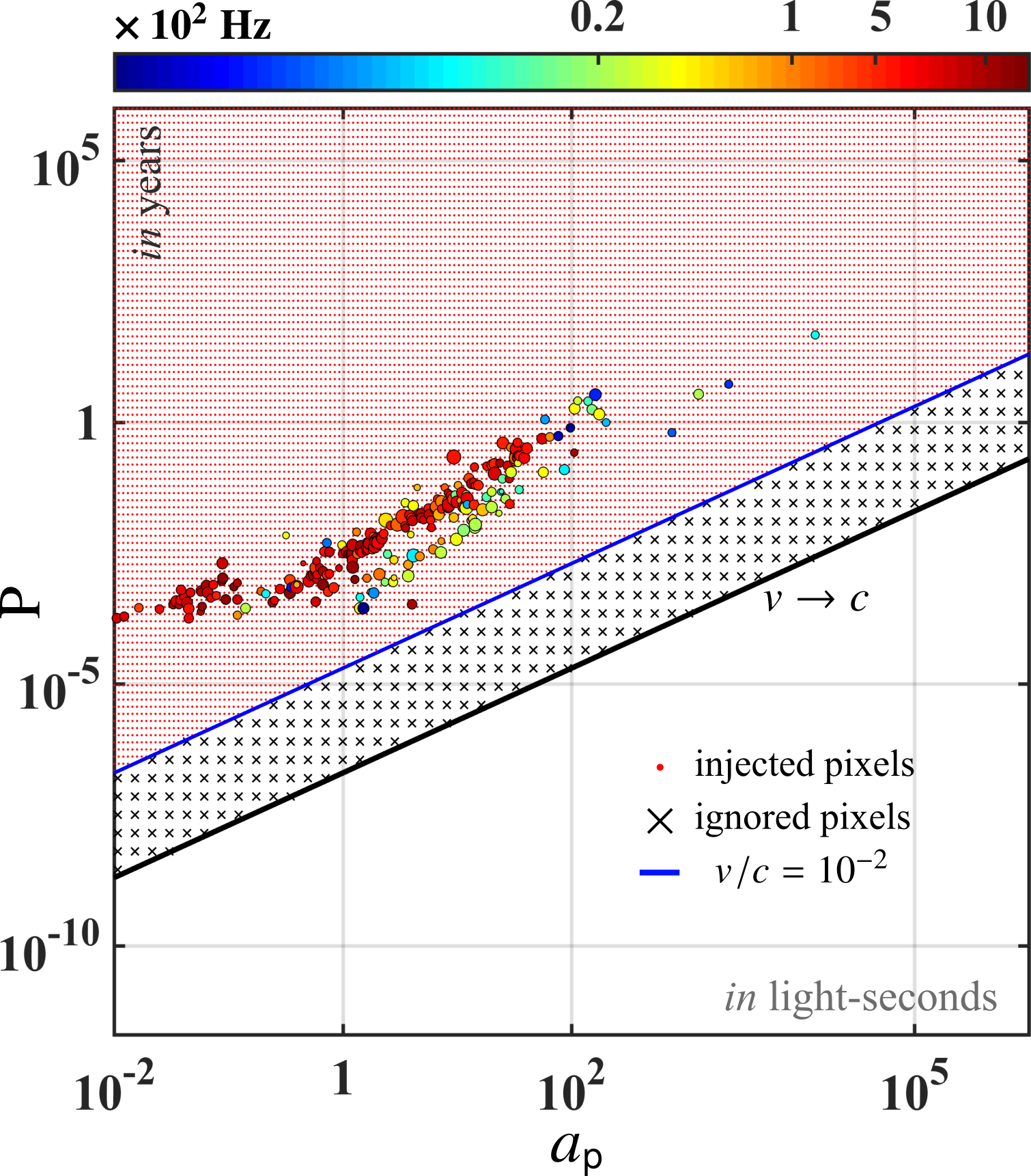}
\caption{{\small The binary waveform parameter space considered in our simulations. Pixels marked with a $\times$ represent parameter space points that were not probed. Known objects from the ATNF catalogue are over-plotted and for these data points, color encodes the intrinsic rotational frequency of the source object while their sizes encode their logarithmic distances ($\propto\mathrm{log}_{10}\mathrm{d}$) from the {\ssb}. The range of distances subtended by these set of known objects from the {\ssb} is ($0.15,\,25$] kpc.}}
\label{fig:pixels}
\end{figure}

In order to record the response of an isolated search to nearly circular binary systems, we simulate binary signals in Gaussian noise. The binary parameters cover the entire physical binary parameter space shown in figure \ref{fig:pixels}. On this simulated data, we perform typical semi-coherent CW searches where the data is partitioned in segments, each spanning the same length in time $\vams{T}{coh}$, which is typically in the range of a few hours to a few days. Then coherent multi-detector searches are performed on each segment separately and the results are combined by appropriately summing the detection statistic values across the segments, one per segment.
{\noindent}We considered nearly 13,000 $(\vars{a}{p},\vams{P}{b})$ templates, as shown in figure \ref{fig:pixels}. The templates marked with $\times$ were ignored since they correspond to values of $\vars{a}{p}$ and $\vams{P}{b}$ that represent extreme physical systems outside our scope of interest, characterized by $v/c > 10^{-2}$. The maximum value of $\vams{P}{b} \sim 3\times10^8\;\mathrm{years}$ is chosen equivalent to the galactic orbital period. The value of the remaining binary parameters is set such that $\vars{\tau}{p} = \vars{\tau}{asc}$, or equivalently $\omega = 0$. In this case, $\vars{\tau}{p}$ encodes a reference time for the phase model and it is set equal to the reference time $\vars{\tau}{ref}$ of the isolated CW search (e.g. see table \ref{table:setup}). Note that the choice of setting $\omega = 0$ is redundant only when the orbital period is much smaller than the coherent observation time ($\vams{P}{b} \ll \vams{T}{coh}$); when $\vams{P}{b} \gtrsim \vams{T}{coh}$, the choice of $\vars{\tau}{p}$ is non-trivial. By setting $\vars{\tau}{p} = \vars{\tau}{ref}$\footnote{In typical CW searches, $\vars{\tau}{ref}$ is usually set at the middle of total observation duration, i.e. $\vars{\tau}{ref} \equiv (\vars{\tau}{start} + \vars{\tau}{end})/2$.}, we are sensitive to maximum Doppler modulation during the total observation duration when $\vams{P}{b} \gtrsim \vams{T}{coh}$. Thus, our choice of setting $\omega = 0$ is in fact least constraining and it encompasses a broad set of general cases without requiring $\omega$ as an additional third variable besides $\vars{a}{p}$ and $\vams{P}{b}$.
\subsubsection*{\centering \tit{Intrinsic parameters of the simulated signals}}
\label{sec:injparams}
The intrinsic signal parameters (denoted by $\uplambda_i$ $=\{f,\dt{f},\alpha,\delta\}$) of the simulated signals are chosen such that
\begin{equation}
\uplambda_i=
\begin{cases}
\;\;\hfil21.25\text{\Hz} < f < 21.30\text{\Hz},\\
\;\;\hfil-2.6\times10^{-9}\text{\Hz/s} < \dt{f} < 2.6\times10^{-10}\text{\Hz/s},\\
\;\;\hfil\alpha = 19.5^\circ, \delta = 57.9^\circ,\\
\;\;\hfil\varr{h}{o} = 3.5\times 10^{-23}.
\label{eq:injranges}
\end{cases}
\end{equation}
These intrinsic signal parameters are chosen to represent a generic case and the results are analytically scalable in $f$ and $\dt{f}$. Citing practical reasons, we only consider a single position in the sky that is neither at the ecliptic poles nor at the equator\footnote{Note that depending on the proximity of the simulated injections to the poles and/or equator on the ecliptic sphere, a small offset will be present in the results. We provide a brief explanation of this effect later in section \ref{sec:skycor}.}. Our choice here corresponds to $(45^\circ,\,45^\circ)$ on the ecliptic sphere; refer to \citep{AdCl} for equations projecting the equatorial coordinates on the ecliptic sphere and vice-versa. The simulated Gaussian noise data is generated at the sensitivity of advanced LIGO during its {\oone} run. The corresponding amplitude spectral density at 21.25{\Hz} is $\sqrt{\mathsf{S_h}}=1.4\times 10^{-22}\text{\Hz}^{-1/2}$. The amplitude $\varr{h}{o}$ quoted in \eqref{eq:injranges} corresponds to a \tit{depth} $\mathcal{D}$ ($= \sqrt{\mathsf{S_h}}/\varr{h}{o}$) of $4.0\text{\Hz}^{-1/2}$. We call this injection and recovery set the \tit{binary-isolated-search} and denote it by a subscript `\textsf{bi}'. In order to calculate the relative loss of {\SNR} for signals due to unaccounted binary modulations in the binary-isolated-search setup, a matching set of \tit{purely} isolated signals are also injected in simulated Gaussian noise and recovered with a \tit{perfectly-matched-search}. This is equivalent to turning the binary orbital parameters off and preserving only the intrinsic signal parameters of the sources, and then recovering the injected signals with a perfectly-matched isolated search\footnote{The `perfect matching' implies that the exact known injected signal parameters are used as a template for the search.}. The perfectly-matched-search injection and recovery simulation extracts the maximum possible value of the detection statistic. Both searches yield $\TwoF$ as the detection statistic for each recovery instance ($\vars{\TwoF}{bi}$ and $\vars{\TwoF}{max}$ respectively) quantifying the relative loss of statistical significance of a template in presence of a binary companion versus in complete isolation. 
\subsubsection*{\centering \tit{Search setup}}
\label{sec:setup}
In order to quantify the effects of $\vams{T}{coh}$ as discussed in the section \ref{sec:tcoh}, we use two search setups with significantly different $\vams{T}{coh}$ values (see table \ref{table:setup}). We refer to them as \tit{short}$-\Tcoh$ \tit{search setup} and \tit{long}$-\Tcoh$ \tit{search setup}. The setups for both these searches are chosen to mimic typical all-sky \EatH semi-coherent searches, e.g. \citep{O1AS20-100, S6BucketStage0}. The details of the setup are given in table \ref{table:setup}. Both searches operate on data containing simulated signals in Gaussian noise with similar total observation lengths in excess of 100 days. In order to carry out the search in a reasonable amount of time, the final resolution in parameter space is approached in two stages, differing by a factor of 10 in frequency and spin-down while keeping the sky-resolution constant. The numbers quoted in table \ref{table:setup} represent the final stage. The frequency bandwidth searched around each injection (denoted by $\Box f$) is large enough to account for maximum drift in frequency due to all possible modulations. In sky, a $10\,\times\,10$ grid patch is sufficient, where each pixel measures $\vams{d}{sky}\times\vams{d}{sky}$ on the projected ecliptic plane with unit dimensionless radius.
\begin{table}[H]
\begin{center}
\bgroup
\def\arraystretch{1.2}
\begin{tabular}{|l|l|l|}
\hline
\hline
\textbf{Quantity} & \textbf{\tit{short}$-\Tcoh$ \tit{setup}} & \textbf{\tit{long}$-\Tcoh$ \tit{setup}}\\
\hline
\hline
$\mathrm{T}_{\mathsf{coh}}$ (hours) & 60 & 210 \\ \hline
$\mathrm{N}_{\mathsf{seg}}$ & 64 & 12       \\ \hline
$\mathrm{T}_{\mathsf{obs}}$ (hours) & $60\times64$ & $210\times12$       \\ \hline
$\delta\!f$ (Hz) & $6.7\times 10^{-7}$ & $3.3\times 10^{-7}$\\ \hline
$\delta\!\dot{f}$ (Hz/s) & $1.3\times 10^{-13}$ & $1.6\times 10^{-13}$ \\  \hline
$\mathrm{d}_\mathsf{sky}$ &  0.024 & 0.068\\
\hline
\hline
\end{tabular}
\egroup
\end{center}
\caption{The short$-\Tcoh$ and long$-\Tcoh$ search setups.}
\label{table:setup}
\end{table}

\subsubsection*{\centering \tit{Results}}
\label{sec:results}
We measure the response of isolated searches to binary signals and the corresponding loss in {\SNR} with the quantity
\begin{equation}
\begin{multlined}
\mu = \frac{\TwoF_\mathsf{max} - \TwoF_\mathsf{bi}}{\TwoF_\mathsf{max} - 4},\label{eq:m}
\end{multlined}
\end{equation}
where $\TwoF$ is the highest detection statistic value in each search. In general, the expected value, denoted by $\langle\TwoF\rangle$, of the maximum of the detection statistic follows $\langle\TwoF_\mathsf{max}\rangle \geq \TwoF_\mathsf{bi} \geq 4$  \citep{FStatSchutz}. This constrains the expected value of $\mu$ to $0\leq \langle\mu\rangle \leq 1$, although a specific realisation can be lower than 0 or greater than 1. In this paper, with a slight abuse of language, we will refer to $\mu$ as the `mismatch', even though the mismatch is a property of a template bank, usually defined in the absence of noise and strictly lies between between 0 and 1. 

We define the shifts in the intrinsic parameters of the loudest candidates from their true positions by
\begin{equation}
\Delta\uplambda_i = 
\begin{cases}
\;\;\hfil\Delta f     = |f_\mathsf{max} - f_\mathsf{bi}|,\\
\;\;\hfil\Delta\dt{f} = |\dt{f}_\mathsf{max} - \dt{f}_\mathsf{bi}|,\\
\;\;\hfil\Delta\mathrm{R} = |\sqrt{(x_\mathsf{max} - x_\mathsf{bi})^2 + (y_\mathsf{max} - y_\mathsf{bi})^2}|,\\
\label{eq:deltal}
\end{cases}
\end{equation}
where, $(x, y) \in [-1, 1]$ are the projected coordinates on the uniform grid in the ecliptic plane\footnote{Refer to equations (14), (15) in \citep{AdCl} for full definition of the projection from ($\alpha, \delta$) in equatorial coordinates to ($\uplambda, \upbeta$) on the ecliptic sphere to ($x, y$) on the projected ecliptic plane. Note that the choice of sky-position for the injection set as described in \eqref{eq:injranges} corresponds to $x = y = 1/2$.}. Hence, $\Delta\mathrm{R}$ is the Euclidean distance in the projected ecliptic plane between the signal's real and apparent position. The four quantities defined in \eqref{eq:m} and \eqref{eq:deltal} are plotted in figure \ref{fig:panel1}; the top row depicts the long$-\vams{T}{coh}$ search and the bottom row shows the short$-\vams{T}{coh}$ search for comparison. We note that $\vams{T}{coh}$ determines the overall scale of the loss of {\SNR}. Moreover, the morphology of the ${f,\dt{f}}$ parameter shifts in the $\vars{a}{p}-\vams{P}{b}$ plane by and large follow the predictions made in section \ref{sec:lowe}. In particular, we indeed find that an isolated search is able to accommodate binary modulations and retain sensitivity to a binary signal in the $\vars{a}{p}-\vams{P}{b}$ plane as dictated by the spin-down characteristics (white line), whereas the magnitude of such an accommodation is especially accounted for by shifts in frequency (red line). This amounts to saying that the time-independent component of the binary modulation is accommodated by a shift in frequency while the time-variation in such a shift in frequency is accommodated by a shift in spin-down; this assertion is also deducible from the discussions within section \ref{sec:lowe}. However, there are also some departures from the analytic estimates; most notably, the frequency shift shows an offset from the expectation and its ``echo'' is correlated to shift patterns in the remaining parameters. It is also interesting to note that the spin-down and sky-position shifts appear to be correlated. We discuss these effects in detail in the next section.

\subsubsection*{\centering \tit{The offset in frequency shifts}}
\label{sec:skycor}
The absolute loss of {\SNR} between the binary-isolated-search and the perfectly-matched-search is determined not only by the phase mismatch due to the binary modulation, but also by the intrinsic `template mismatch' of the grid-based binary-isolated-search. The quantities depicted in figure \ref{fig:panel1} contain the combined outcome of both these effects. The template mismatch leads to a small variance in the estimated signal parameters. The template mismatch occurs because the demodulation for Earth's relative motion to the {\ssb} is performed at the observed sky-location instead of the true unknown sky-location, thus leaving a small residue due to this error in sky localisation. In the end, the final outcome shown in figure \ref{fig:panel1} is a result of the residual modulation due to template mismatch added on top of the binary modulation.

In order to quantify the background frequency offset due to template mismatch, we begin by writing the expression for residual modulation due to an error in estimated sky-position. This relation is calculated following \eqref{eq:dopred}--\eqref{eq:c1f} and the result is a function of the error in sky-localization on the projected ecliptic plane (denoted by $\delta \mathrm{R}$), such that
\begin{equation}
\begin{gathered}
{\langle\Delta\vars{f}{tem}\rangle} \lesssim \frac{2\uppi f\vars{a}{orb}}{\vams{P}{orb}} \delta \mathrm{R} \sim 1.0\times10^{-4}f\delta \mathrm{R},\\
\langle\Delta\vars{\dt{f}}{tem}\rangle \lesssim \frac{4\uppi^2 f\vars{a}{orb}}{\vams{P}{orb}^2} \delta \mathrm{R} \sim 2.0\times10^{-11}f\delta \mathrm{R}.
\end{gathered}\label{eq:t1f}
\end{equation}
where $\vams{P}{orb}$ and $\vars{a}{orb}$  are the orbital period and the projected semi-major axis of Earth's orbit around the {\ssb} respectively\footnote{We have ignored the effects Earth's spin since $\langle \vars{v}{spin}\rangle/\langle \vars{v}{orb}\rangle \ll 1$.}. We test this constraint by adding a template grid to the perfectly-matched-search, thereby including the effects of template mismatch; we refer to this search by \tit{mismatched-search}. On performing the mismatched-search on the long$-\Tcoh$ search and short$-\Tcoh$ search setups, the observed shift in the parameter space due to template mismatch averaged over 13,000 simulations is shown in table \ref{table:shifts} relative to the respective grid spacings. The constraints defined in \eqref{eq:t1f} are also stated alongside and we find that they are respected. The difference in departure of the observed values from the expected values between the two columns in table \ref{table:shifts} is due to different coherent lengths and total observation times of the setups. We note that the shift in sky-position which results in the said template mismatch is dependent on the true signal location in the sky through $\delta \mathrm{R}\propto \ccos \upbeta$, i.e. for sources near the ecliptic poles ($\upbeta = \pm 90^\circ$), $\delta \mathrm{R} \rightarrow 0$, while for sources near the ecliptic equator ($\upbeta = 0^\circ$), $\delta \mathrm{R}$ approaches its maximum value.
\end{multicols}
\begin{center}
\begin{sideways}
\begin{minipage}{245mm}
{\textbf{
\hspace{60pt}\tit{mismatch}       
\hspace{110pt}\tit{frequency shift}
\hspace{110pt}\tit{spin-down shift}
\hspace{100pt}\tit{shift in sky}}  
\hspace{70pt}\par\medskip}
\includegraphics[width=60.5mm]{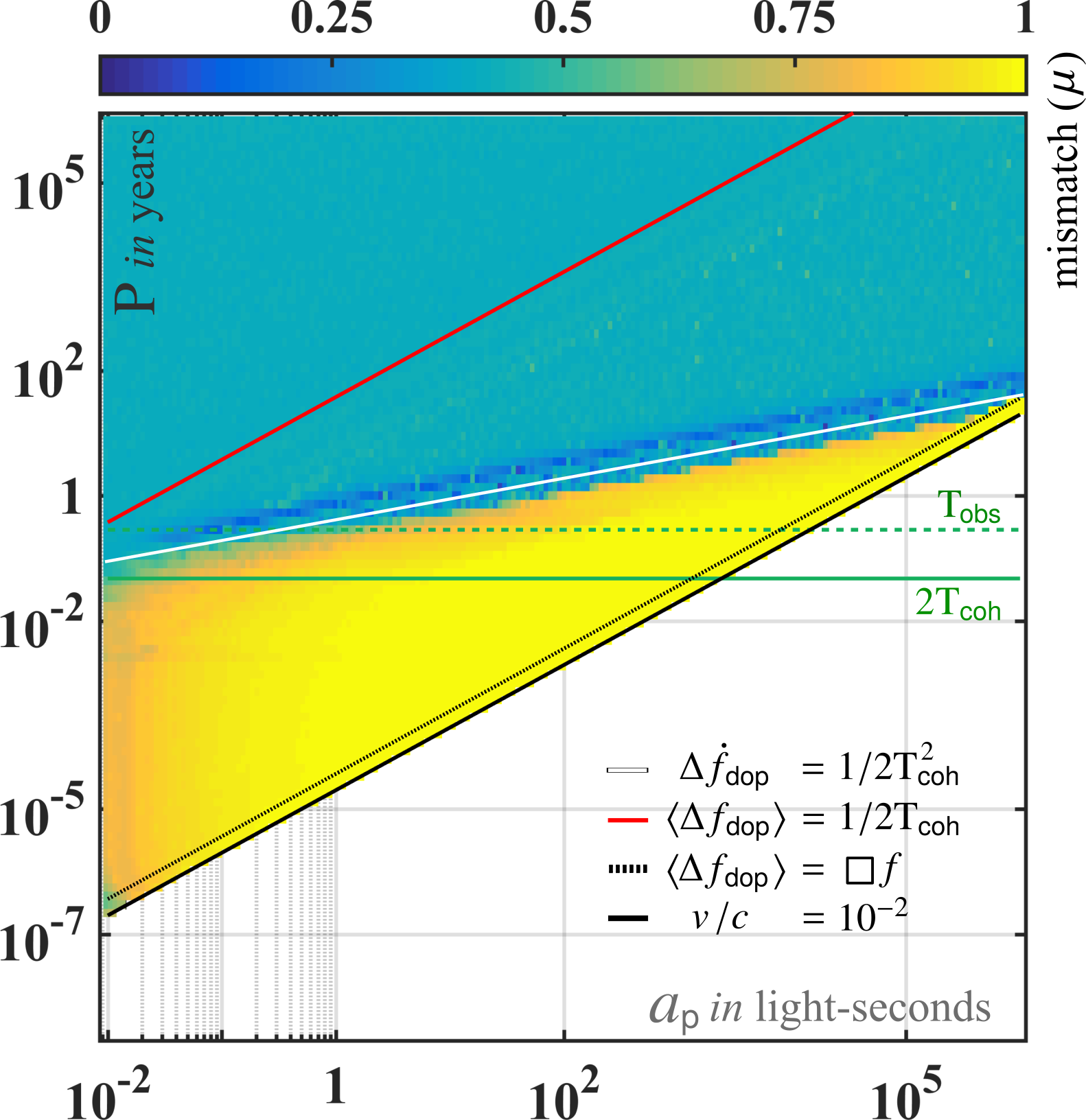}
\includegraphics[width=60.5mm]{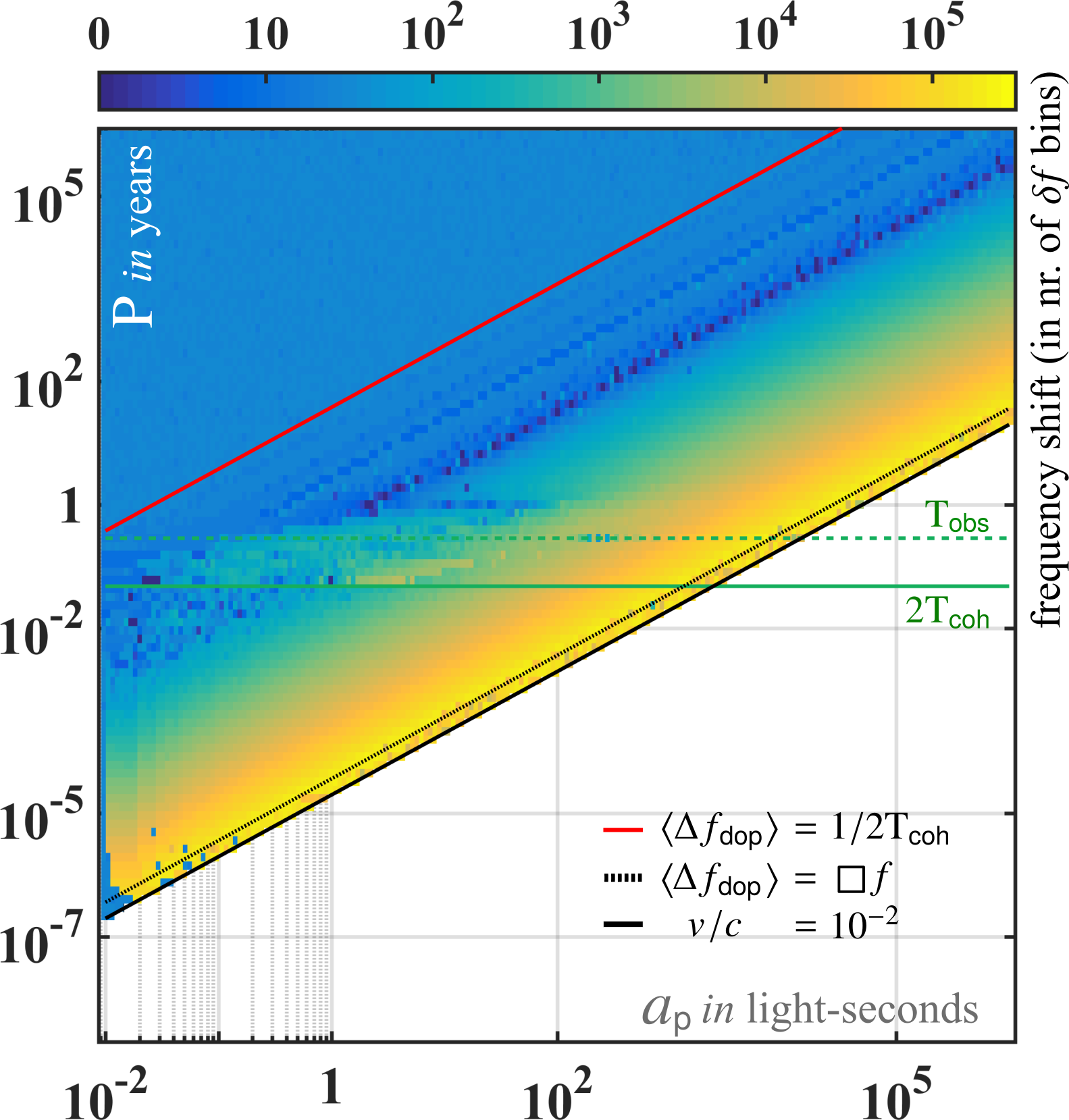}
\includegraphics[width=60.5mm]{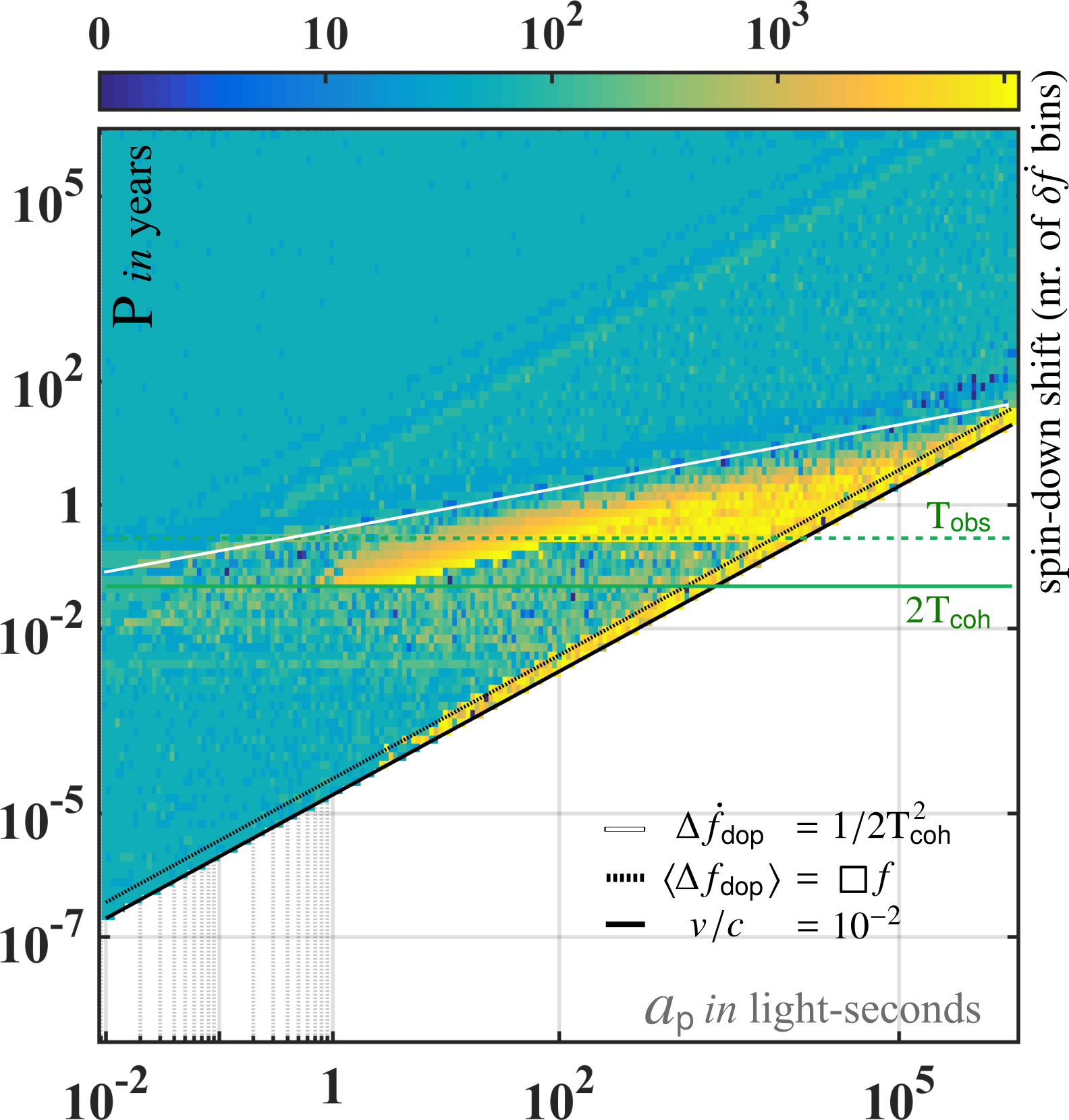}
\includegraphics[width=60.5mm]{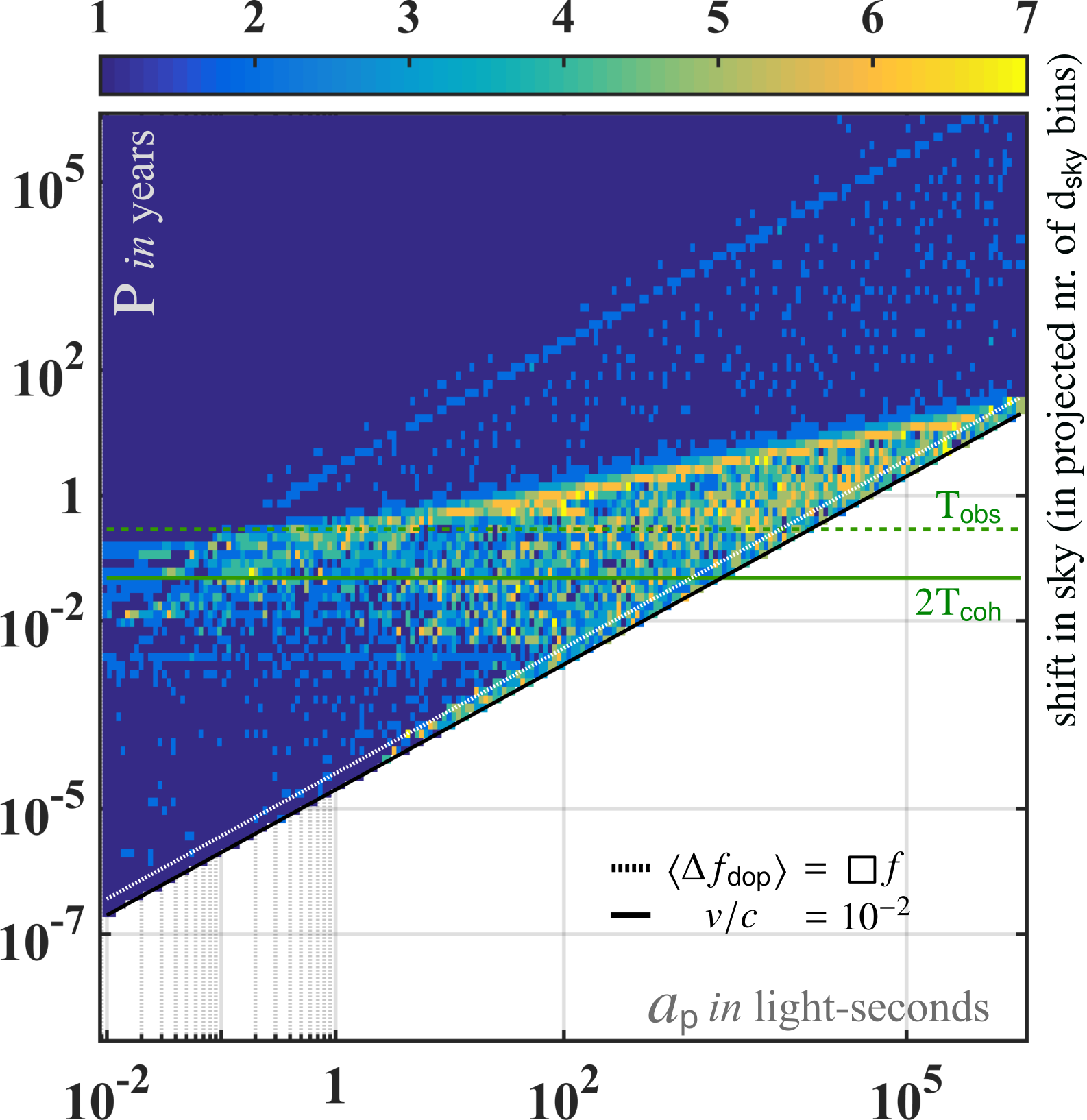}
\par\medskip
\includegraphics[width=60.5mm]{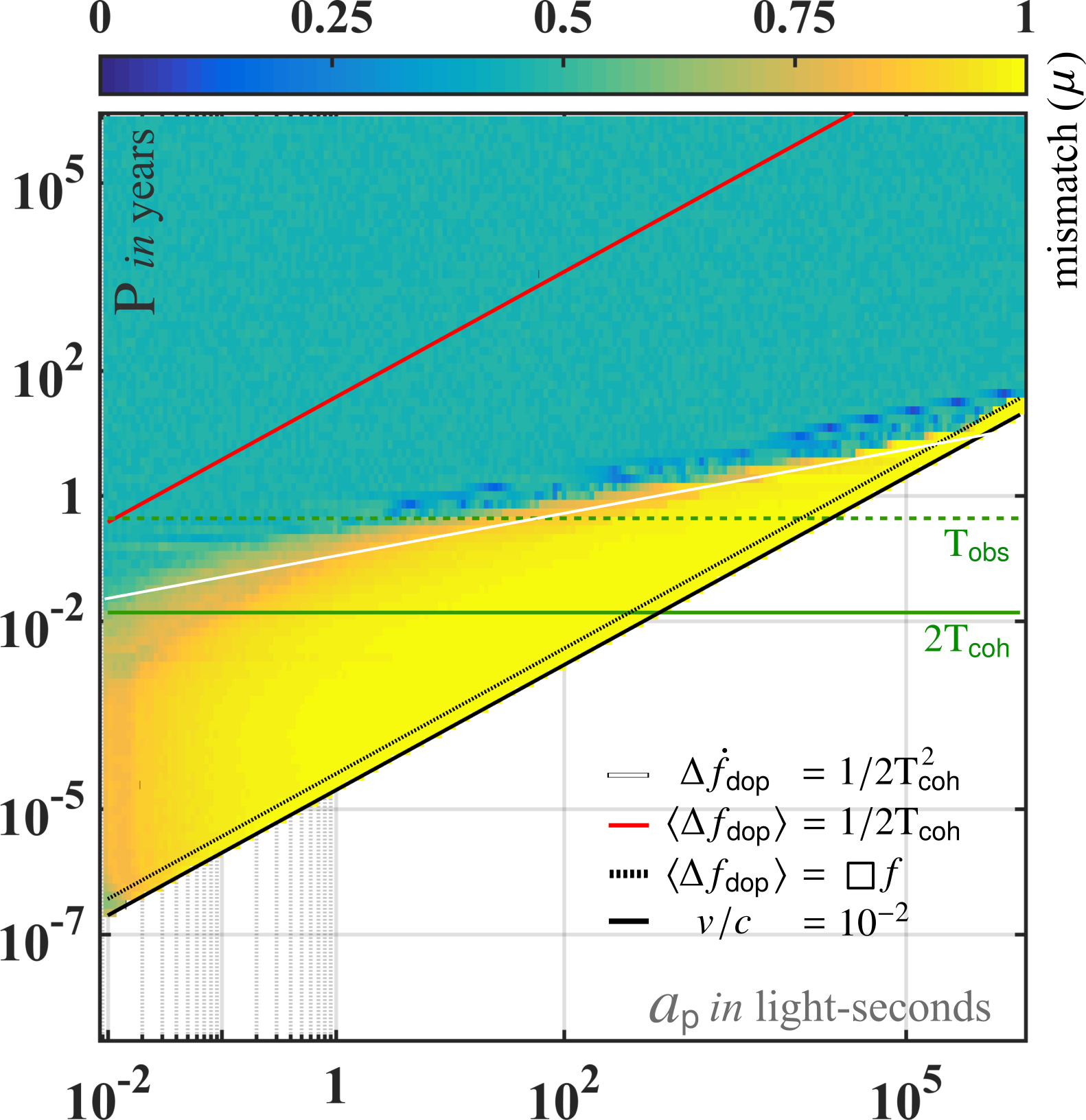}
\includegraphics[width=60.5mm]{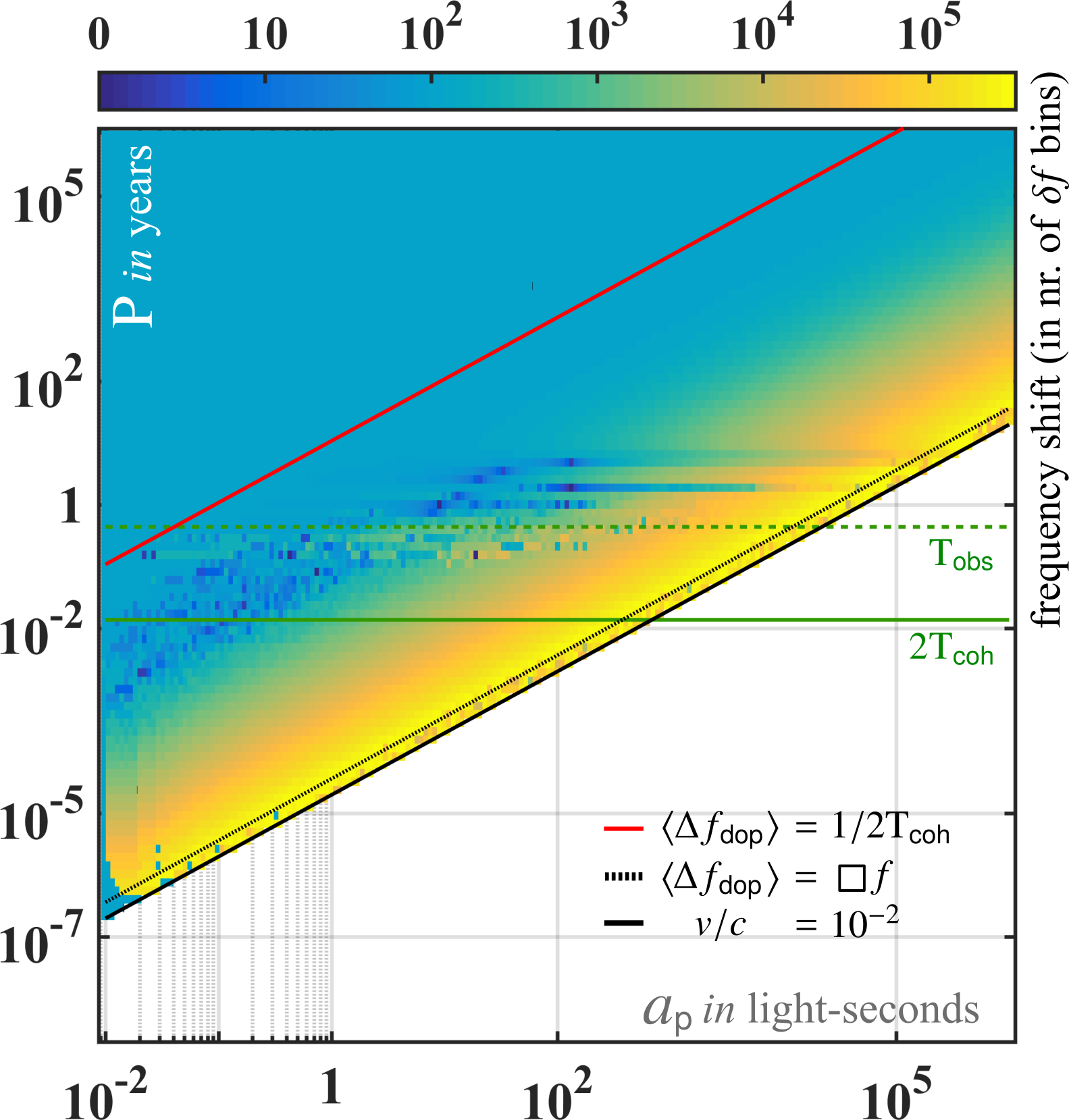}
\includegraphics[width=60.5mm]{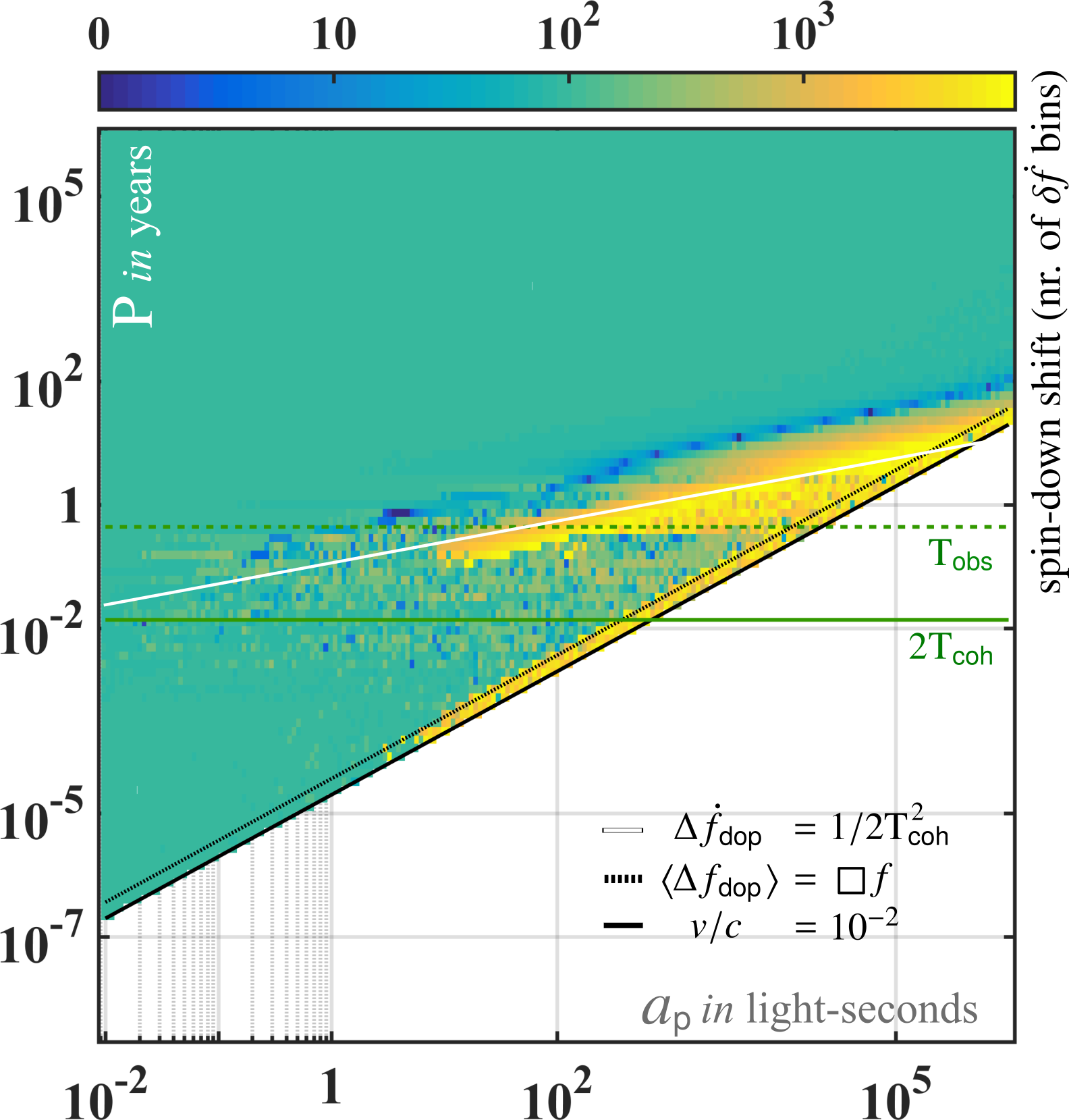}
\includegraphics[width=60.5mm]{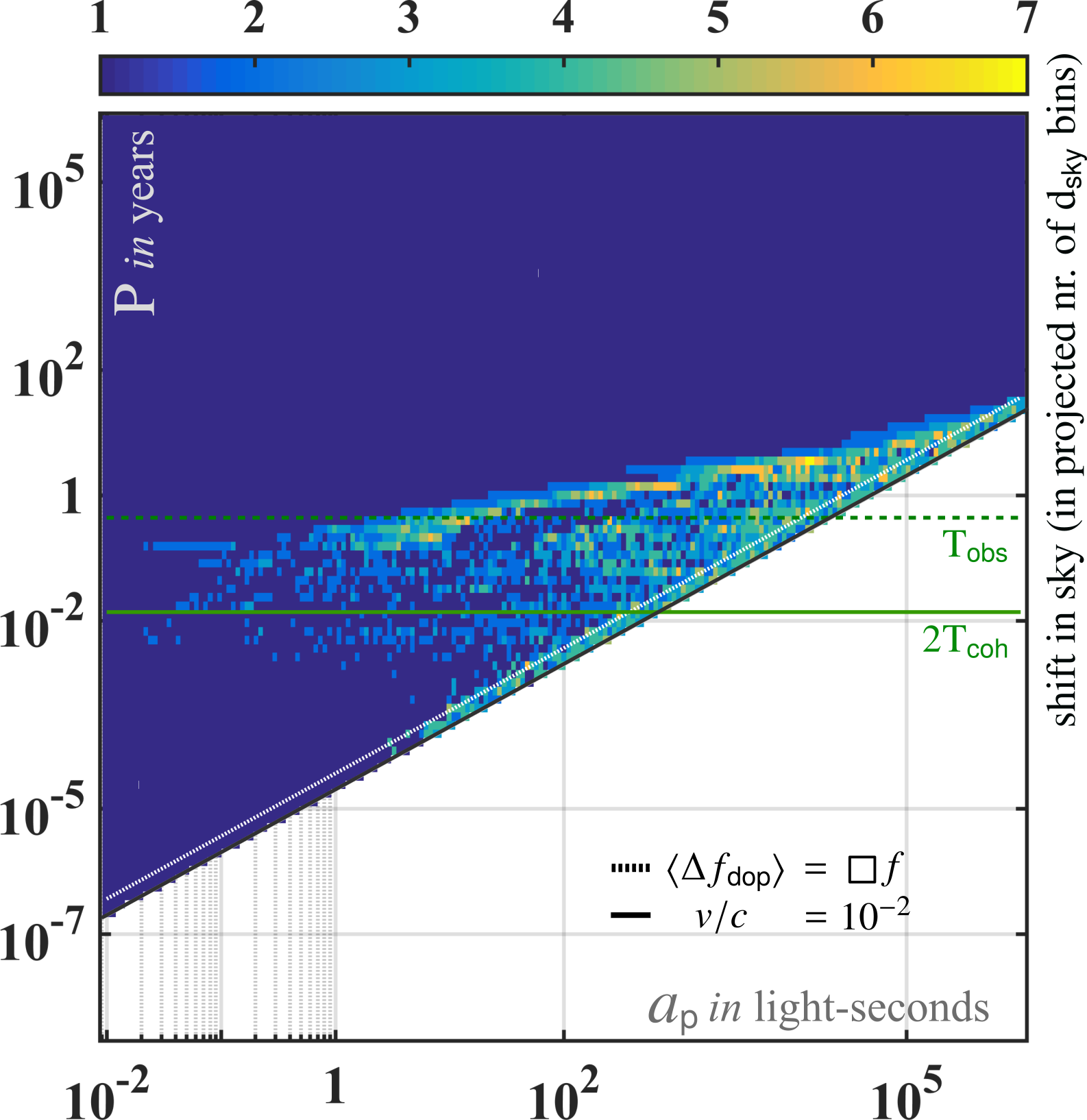}
\captionof{figure}{\small \tit{Absolute} loss of {\SNR}: \tit{first column}: observed mismatch as a function of the binary parameters ($\mu$); \tit{second column}: observed shift in frequency of the signal as a function of the binary parameters (in number of $\delta\! f$ bins); \tit{third column}: observed shift in spin-down of the signal as a function of the binary parameters (in number of $\delta\!\dot{f}$ bins); \tit{fourth column}: observed shift in position of the signal in sky as a function of the binary parameters (in number of projected $\mathrm{d}_\mathsf{sky}$ bins). The \tit{top row} corresponds to the long$-\Tcoh$ search; the \tit{bottom row} shows the results of the short$-\Tcoh$ search setup.}
\label{fig:panel1}
\end{minipage}
\end{sideways}
\end{center}
\pagebreak
\begin{multicols}{2}
{\noindent}In addition to this, the second equation in \eqref{eq:t1f} also encodes the correlation between the observed shifts in spin-down and sky-position in figure \ref{fig:panel1}. A similar correlation is also present between frequency and sky-position (first equation) but it is mostly invisible due to relatively large frequency modulations added on top by the binary companion.
\begin{table}[H]
\begin{center}
\bgroup
\def\arraystretch{1.2}
\begin{tabular}{|l|l|l|}
\hline
\hline
\textbf{Quantity} & \textbf{\tit{short}$-\Tcoh$ \tit{setup}} & \textbf{\tit{long}$-\Tcoh$ \tit{setup}}\\
\hline
\hline
$\delta \mathrm{R}/\vams{d}{sky}$ & $0.6 < 1$ & $0.6 < 1$ \\ \hline
$\langle\Delta\vars{f^\mathsf{obs}}{tem}\rangle/\delta\!f$ & $100 < 150$ & $30 < 360$ \\ \hline
$\langle\Delta\vars{\dt{f}^\mathsf{obs}}{tem}\rangle/\delta\!\dot{f}$ & $90 < 100$ & $59 < 60$ \\ \hline
\hline
\hline
\end{tabular}
\egroup
\end{center}
\caption{Background shifts in parameter space due to template mismatch.}
\label{table:shifts}
\end{table}

\begin{figure}[H] 
\centering\includegraphics[width=78mm]{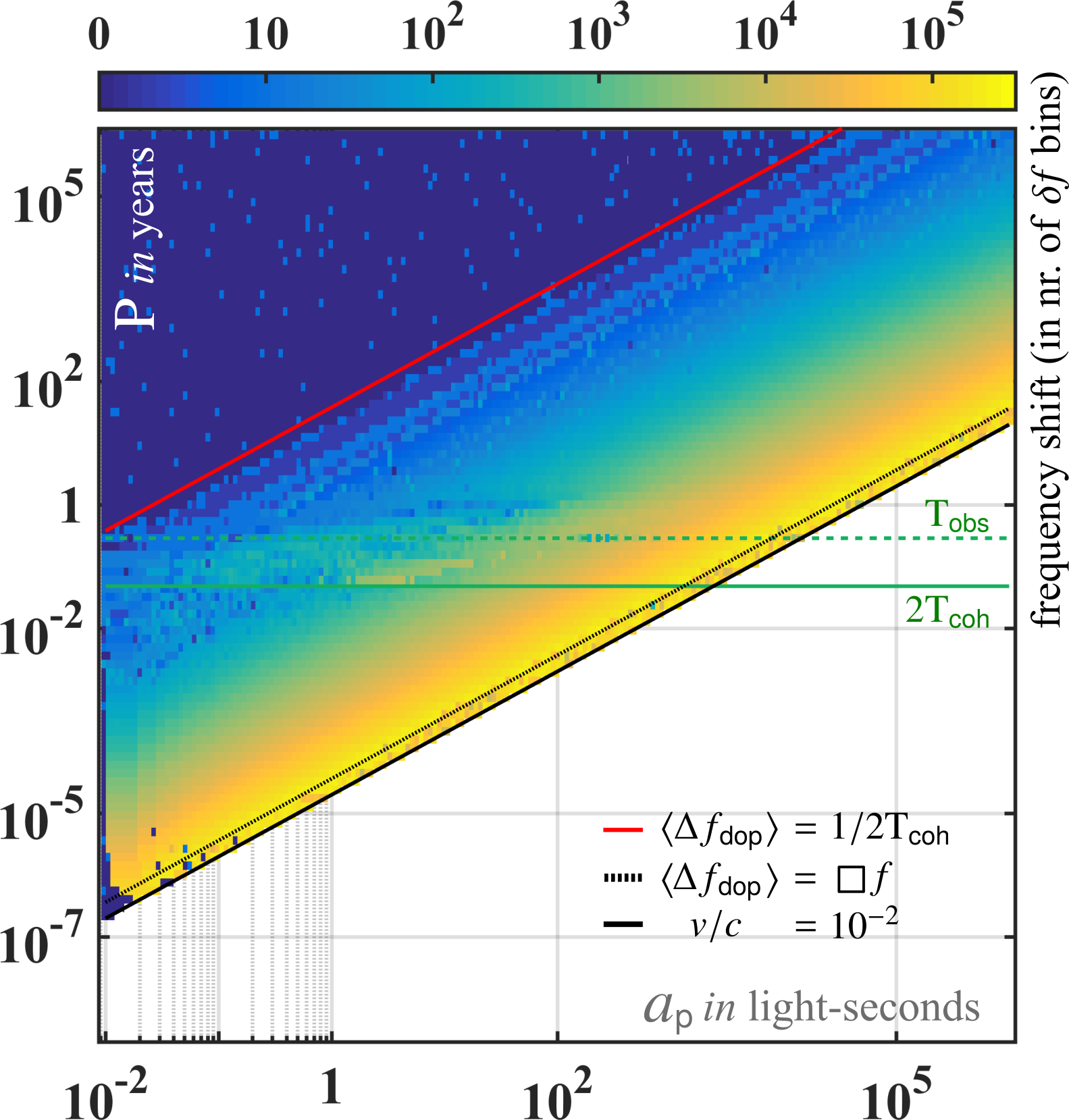}
\caption{{\small Disappearance of offset in frequency shift after subtracting the effects of template mismatch. We note that the characteristics shown above correspond well with the expected modulation in frequency due to the binary companion.}}
\label{fig:tm}
\end{figure}

{\noindent}We further demonstrate this effect in figure \ref{fig:tm} by subtracting the template mismatch from the total shift plotted in figure \ref{fig:panel1} (e.g. the frequency shift panel) corresponding to the long$-\Tcoh$ search setup. We find that post-subtraction, the offset in the observed total shift in the frequency of the signal disappears as expected. 
\section{Conclusions} 
\label{sec:conc}
We have successfully estimated the apparent shift in the intrinsic signal parameters of a CW source in a binary orbit with a purely isolated search and the resulting loss in \textsf{SNR}, by means of a simplified model. We have confirmed the estimates through Monte-Carlo simulations. 

\begin{figure}[H]
\centering\includegraphics[width=78mm]{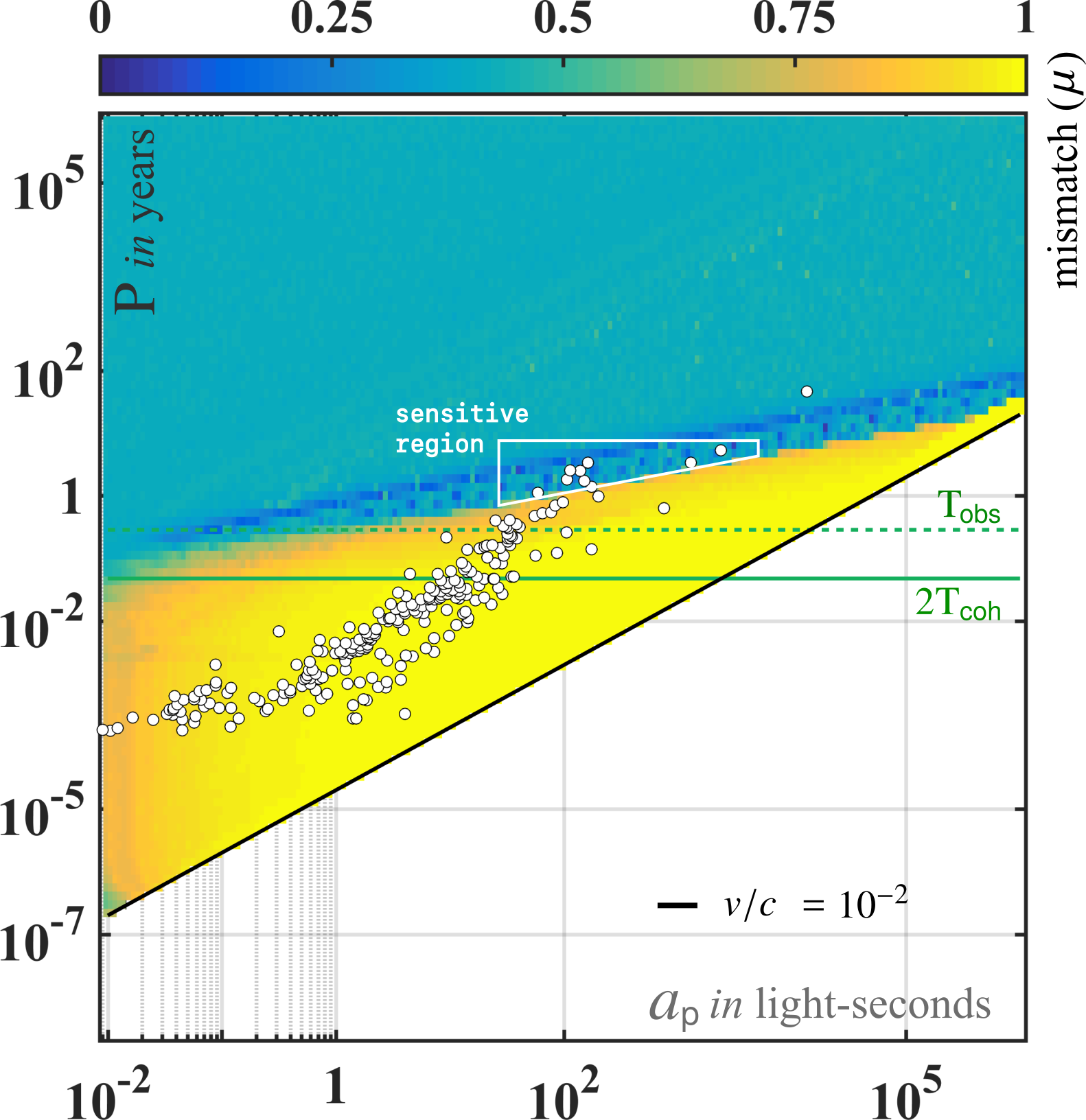}
\caption{{\small $\vars{a}{p}-\vams{P}{b}$ plane overlaying known population of neutron stars in binaries on top of the mismatch values for the long$-\Tcoh$ search setup. Regions of low mismatch values overlapping with the astrophysical population are of interest and an example is shown in white trapezoid.}}
\label{fig:box}
\end{figure}
\begin{figure}[H]
\centering\includegraphics[width=78mm]{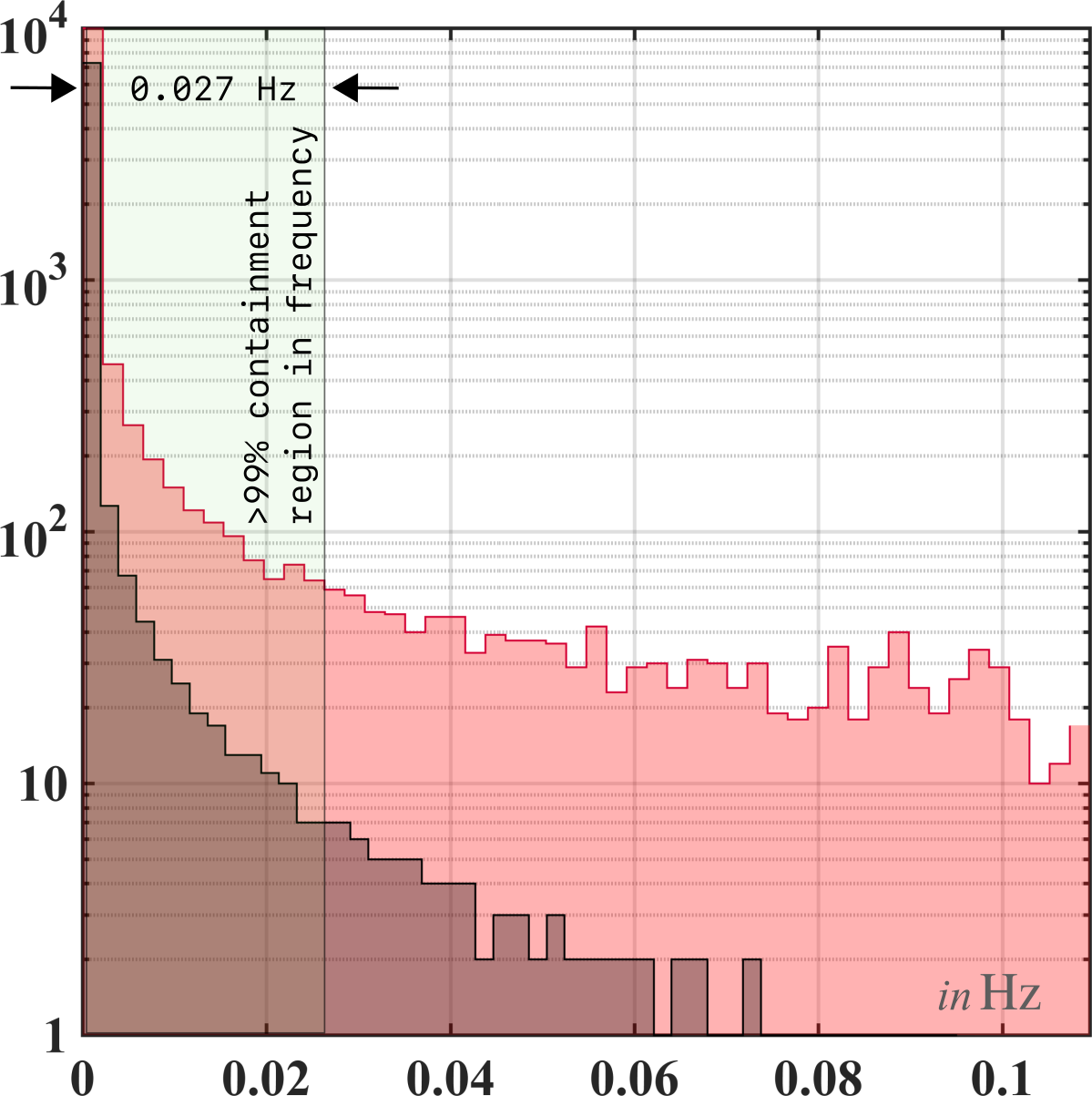}
\caption{{\small Containment region in frequency. The red histogram describes the overall distribution of frequency shifts whereas the gray histogram shows the frequency shifts \tit{only} in the selected region in $\vars{a}{p}-\vams{P}{b}$ plane.}}
\label{fig:dist}
\end{figure}

{\noindent}The agreement between the analytic estimates and the simulation results suggests that for future searches a \tit{macroscopic} assessment of the mismatch characteristics does not require expensive simulations but that such characteristics can be analytically estimated. These estimates assist the targeted binary follow-up searches of candidates from an isolated search. 

To illustrate this case, we assume a hypothetical candidate from long$-\Tcoh$ search at 20{\Hz}. In order to determine whether this candidate harbours a possible binary companion, we must ascertain whether the long$-\Tcoh$ search is sensitive in an astrophysically accessible region in the $a-\vams{P}{b}$ plane. Our results show that even a long time baseline semi-coherent search, such as with $\Tcoh = 210$ hours, is sensitive to signals that are not approximately isolated, as shown in figure \ref{fig:tcoh} (top panel). In regions of astrophysical interest where the search sensitivity is retained to possible binary modulations, we extract the expected shifts in the intrinsic signal parameters according to figure \ref{fig:panel1}. For instance, neutron stars in binary orbits with brown dwarfs and similar objects could be interesting candidates; see figure \ref{fig:tcoh} (bottom panel). One such selected region in $a-\vams{P}{b}$ plane in shown in figure \ref{fig:box} and described by
\begin{equation}
\begin{gathered}
a \in (30, 5000)\;\text{light--seconds},\\
\vams{P}{b} \in (0.5, 7)\;\text{years},\\
\mu \lesssim 0.5.
\end{gathered}\label{eq:bin}
\end{equation}
{\noindent}The distributions of the corresponding shifts in intrinsic signal parameters within this region provide us with information about the regions of maximal \tit{containment} (e.g. $>99\%$), i.e. ranges in frequency, spin-down and sky-position in which the true signal parameters must lie. In figure \ref{fig:dist}, we show the distribution of frequency shifts and the calculated containment in frequency for the region selected in figure \ref{fig:box}. In a similar fashion, containment regions in spin-down and sky-position are also calculated, and the values are given by,
\begin{equation}
\uplambda_\mathsf{box}^\mathsf{long-\Tcoh\;search}=
\begin{cases}
\;\;\hfil\Delta f \sim 0.027\text{\Hz},\\
\;\;\hfil\Delta\dt{f} \sim 1.3\times 10^{-11}\text{\Hz/s},\\
\;\;\hfil\Delta\mathrm{R} \sim 16\,\vams{d}{sky}.
\label{eq:boxes}
\end{cases} 
\end{equation}
{\noindent}In these significantly reduced containment regions in the intrinsic parameter space chosen from astrophysically motivated priors in the $a-\vams{P}{b}$ plane, dedicated binary searches can perform significantly sensitive follow-up searches of candidates from an isolated search.

\section{Acknowledgments} 
\label{sec:ack}
The simulations for this study were performed on the ATLAS cluster at MPIG Hannover. We thank Bruce Allen for allowing us to this wonderful computing facility.
\bibliographystyle{plainnat}
\bibliography{Bibliography}
\end{multicols}
\end{document}